\documentclass[a4paper,11pt]{article}
\pdfoutput=1 

\usepackage[utf8]{inputenc}

\usepackage{jcappub} 

\usepackage[T1]{fontenc} 

\usepackage{amsmath}
\usepackage{amssymb}
\usepackage{graphicx}
\usepackage{hyperref}
\usepackage{color}
\usepackage{commath}
\usepackage{url}
\usepackage{bm}
\usepackage{braket}
\usepackage{tikz}
\usetikzlibrary{positioning}
\def\layersep{2.5cm}
\usepackage{url}
\usepackage{accents}
\usepackage{natbib}
\usepackage{subfigure}

\hypersetup{colorlinks=true,citecolor=blue, linkcolor=red, urlcolor=blue}

\newcommand{\dd}{\mathrm{d}}

\title{\boldmath Neural network reconstruction of scalar-tensor cosmology}

\author[a,b]{Konstantinos F. Dialektopoulos,}
\author[c]{Purba Mukherjee,} 
\author[d,e]{Jackson Levi Said,}
\author[e]{Jurgen Mifsud}

\affiliation[a]{Department of Mathematics and Computer Science, Transilvania University of Brasov, Eroilor 29, Brasov, Romania}
\affiliation[b]{Laboratory of Physics, Faculty of Engineering, Aristotle University of Thessaloniki, 54124 Thessaloniki, Greece}
\affiliation[c]{Physics and Applied Mathematics Unit, Indian Statistical Institute, Kolkata - 700108, India}
\affiliation[d]{Institute of Space Sciences and Astronomy, University of Malta, Malta, MSD 2080.}
\affiliation[e]{Department of Physics, University of Malta, Malta, MSD 2080.}

\emailAdd{kdialekt@gmail.com}
\emailAdd{purba16@gmail.com}
\emailAdd{jackson.said@um.edu.mt}
\emailAdd{jurgen.mifsud@um.edu.mt}

\abstract{Neural networks have shown great promise in providing a data-first approach to exploring new physics. In this work, we use the full implementation of late time cosmological data to reconstruct a number of scalar-tensor cosmological models within the context of neural network systems. In this pipeline, we incorporate covariances in the data in the neural network training algorithm, rather than a likelihood which is the approach taken in Markov chain Monte Carlo analyses. For general subclasses of classic scalar-tensor models, we find stricter bounds on functional models which may help in the understanding of which models are observationally viable.}

\keywords{cosmology, reconstruction, horndeski gravity, neural networks}

\begin{document}
\maketitle

\section{Introduction} \label{sec:intro}

$\Lambda$CDM model is universally accepted as the concordance model of cosmology both for astrophysical and cosmological regimes \cite{Peebles:2002gy,Copeland:2006wr}. In this setting, cold dark matter (CDM) acts on galactic scales \cite{Baudis:2016qwx,XENON:2018voc} while the accelerating expansion of the Universe \cite{Riess:1998cb,Perlmutter:1998np} is described by a cosmological constant in the Einstein-Hilbert action \cite{Mukhanov:991646}. Together with cosmic inflation \cite{Guth:1980zm,Linde:1981mu}, this gives the standard model of cosmology. However, problematic elements remain in the model ranging from theoretical issues with the cosmological constant \cite{Weinberg:1988cp}, as well as its UV completeness \cite{Addazi:2021xuf} and other open questions such as the direct observation of dark matter \cite{LUX:2016ggv,Gaitskell:2004gd}. More recently, a new feature of the theory has gained prominence in which the predictions of the Hubble constant turn out to be in tension between certain different surveys \cite{Abdalla:2022yfr}. The Hubble tension has prompted a reevaluation of both the gravitational foundations of the standard cosmological model \cite{CANTATA:2021ktz}, as well as many other considerations \cite{DiValentino:2020vhf,DiValentino:2020zio,DiValentino:2020vvd,Staicova:2021ajb,DiValentino:2021izs,Perivolaropoulos:2021jda,DiValentino:2022oon,SajjadAthar:2021prg}.

The Hubble tension has drastically increased the reporting of Hubble constant values from different phenomena showing a growing discrepancy between direct and indirect measurements of $H_0$, where the latter is reported by assuming a $\Lambda$CDM cosmology \cite{Poulin:2023lkg}. The latest reported values of the Hubble constant from the Planck and ACT collaborations are respectively $H_0^{\rm P18} = 67.4 \pm 0.5 \,{\rm km\, s}^{-1} {\rm Mpc}^{-1}$ \cite{Aghanim:2018eyx} and $H_0^{\rm ACT-DR4} = 67.9 \pm 1.5 \,{\rm km\, s}^{-1} {\rm Mpc}^{-1}$ \cite{ACT:2020gnv}. On the other hand, direct measurements from local sources have produced Hubble constant values from various phenomena. The SH0ES Team has determined the best value of the Hubble constant to be $H_0^{\rm R20} = 73.2 \pm 1.3 \,{\rm km\, s}^{-1} {\rm Mpc}^{-1}$ \cite{Riess:2020fzl} which is based on observations of Supernovae Type Ia (SN-Ia) calibrated with Cepheid stars. In line with this measurement, the strong lensing of quasars has produced $H_0^{\rm HW} = 73.3^{+1.7}_{-1.8} \,{\rm km\, s}^{-1} {\rm Mpc}^{-1}$ which was reported by the H0LiCOW Collaboration \cite{Wong:2019kwg}. Other reported values also give lower Hubble constant values such as that based on the Tip of the Red Giant Branch (TRGB) calibration technique which has been reported to give a value $H_0^{\rm F20} = 69.8 \pm 1.9 \,{\rm km\, s}^{-1} {\rm Mpc}^{-1}$ \cite{Freedman:2020dne}. While each survey is internally consistent, some combinations of surveys are in tension with each other, which is particularly prevalent when the standard model of cosmology is used to make calculations.

The response to the cosmic tensions problem more broadly has been varied. It is exceedingly unlikely that this will be found to be the result of systematics or a feature of observatories. To confront this problem there have been several suggestions for modifications to the standard cosmological model including modifications to early Universe dark energy \cite{Poulin:2023lkg}, the neutrino sector \cite{DiValentino:2021imh}, as well as renewed interest in gravitational models \cite{Addazi:2021xuf,CANTATA:2021ktz,Cai:2019bdh,Ren:2022aeo,Bernardo:2021qhu,Briffa:2020qli,LeviSaid:2021yat}. There have also been interesting studies in which cosmography is used to approach the problem \cite{Bargiacchi:2021fow,Capozziello:2019cav,Bamba:2012cp}. A particularly interesting approach to modifying cosmological models is through the Horndeski gravity \cite{Horndeski:1974wa} where a general scalar-tensor formalism is adopted under the condition that the field equations are second order in metric derivatives. While the generality of these models is interesting for constructing many varied equations of motion in many varied settings, it can pose a problem for determining viable candidates for cosmological scenarios \cite{Zumalacarregui:2013pma,Gleyzes:2014dya,Kobayashi:2019hrl,Bernardo:2021izq,Bernardo:2021bsg}. In this vein, the recent multimessenger observations by the LIGO-Virgo collaboration, namely GW170817 \cite{LIGOScientific:2017vwq}, and Fermi Gamma-ray Burst Monitor, which observed the electromagnetic counterpart GRB170817A \cite{Goldstein:2017mmi}, put strict bounds on the speed of gravitational waves in comparison to the speed of light. This has drastically reduced the model space of Horndeski gravity making it more realistically searchable using numerical methods \cite{Ezquiaga:2018btd}. In this work, we aim to use novel developments in artificial neural network (ANN) \cite{10.2307/j.ctt4cgbdj} reconstruction schemes to build a model selection approach that is data-driven.

Currently, implementations of machine learning in modified cosmological models have mainly taken the form of Gaussian Processes (GP) \cite{10.5555/1162254} which is based on a kernel that characterizes the covariance distribution of a set of data points over some range. These kernels are described by a set of nonphysical so-called hyperparameters which can be fit using traditional methods of Bayesian statistics. There have been numerous works on reconstructing cosmological parameters in this way \cite{Busti:2014aoa,Busti:2014dua,Seikel:2013fda,Bernardo:2021mfs,Yahya:2013xma,2012JCAP...06..036S,Shafieloo:2012ht,Benisty:2020kdt,Benisty:2022psx,Bernardo:2022pyz,Escamilla-Rivera:2021rbe,Bernardo:2021cxi,Mukherjee:2021epjc,Mukherjee:2022pdu}. This approach has also been used to reconstruct certain models of cosmology \cite{Cai:2019bdh,Bernardo:2021qhu,Ren:2022aeo,Briffa:2020qli,LeviSaid:2021yat}. However, GP has several drawbacks that make it problematic primary among which is its overfitting for low values of redshift and an over-reliance on the choice of kernel function which can alter the constraints on the Hubble constant in a significant way for some scenarios.

A new approach to performing reconstructions in modified cosmological models using machine learning applications has been to use ANNs using expansion data. In this setup, artificial neurons are modeled on their biological analog which are then organized into layers through which signals or inputs are mapped to output parameters. This could take the form of redshift inputs and Hubble data outputs \cite{aggarwal2018neural,Wang:2020sxl,Gomez-Vargas:2021zyl}. This system of neurons would contain a huge number of hyperparameters that can be optimized for particular data sets through ANN training. Recently, Ref.~\cite{Wang:2019vxv} presented one such implementation in which Hubble data is produced using ANNs, which was then further studied in Ref.~\cite{Dialektopoulos:2021wde} using a number of null tests. The reason why GP is very attractive as an approach through which to study modified cosmological models is that it naturally produces higher order derivatives of the parameter being reconstructed from data. Given that most general cosmological models contain derivatives of the Hubble parameter, this is very advantageous in these studies. It is more cumbersome to mimic this for ANN systems. On the other hand, one approach was detailed in Ref.~
\cite{Mukherjee:2022yyq} where the $H'(z)$ was indeed reconstructed together with its uncertainties using a Monte Carlo method. This has opened the way for an equivalent approach in which cosmological models are selected using ANNs rather than the increasingly problematic GP approach. Saying that, one drawback of this study is that it can only be applied to Gaussian data. In the current work, we extend this analysis to also include correlated data which is more realistic given the current plethora of cosmological data sets. 

In this paper, we extend the ANN approach for reconstructing observational data by incorporating more realistic complexity in this data during the training of the ANN hyperparameters. In this way, the system more accurately mimics observations being made for the various data sets being used. The work is divided as follows: in Sec.~\ref{sec:ann_intro} we briefly introduce the mechanics of ANN systems, while in Sec.~\ref{sec:ANN_results} we show how this can be used to incorporate more complex data into reconstruction schemes. In Sec.~\ref{sec:model_results} this work is connected to data-driven Horndeski model implementations, while in Sec.~\ref{sec:conclusion} we summarize our main results and give a conclusion.

\section{Horndeski gravity}

Horndeski gravity is the most general scalar-tensor theory of gravity with a single scalar field $\phi$ in four dimensions, that gives second-order field equations both for the metric and for the scalar field. It was introduced in 1974 by G.~W.~Horndeski \cite{Horndeski:1974wa} but for many years it was treated as a mathematical exercise instead of a promising theory of gravity. In the late 2000's it was reintroduced as the generalized covariant galileon model \cite{Nicolis:2008in,Deffayet:2009wt,Deffayet:2009mn}, which coincides with Horndeski's proposed theory. Since then, it has been used in many applications such as self-accelerating cosmological solutions \cite{Silva:2009km}, non-trivial black hole solutions \cite{babichev:hal-01554627}, underlying symmetries \cite{Capozziello:2018gms}, and more. After the observation of GW170817 part of it was severely constrained \cite{Copeland:2018yuh} and there have been several attempts either to extend it \cite{Gleyzes:2014dya,Langlois:2015cwa} or to reformulate it in non-Riemannian geometries \cite{Bahamonde:2019shr,Bahamonde:2022cmz}. Indeed, some of the terms that were eliminated in the standard Horndeski, could revive in its teleparallel analog, i.e. BDLS theory \cite{Bahamonde:2019ipm} and many applications have been studied in the context of this theory \cite{Bahamonde:2020cfv,Bahamonde:2021dqn,Dialektopoulos:2021ryi}.

The action of the theory reads as
\begin{equation}\label{eq:Horndeski-action}
    \mathcal{S} = \int \dd ^4 x \sqrt{-g} \sum _{i=2} ^5 \mathcal{L}_i + \mathcal{S}_{\rm matter}(\psi,g_{\mu\nu})\,.
\end{equation}
We denote all the matter fields collectively as $\psi$ and the Lagrangians terms $\mathcal{L}_i$ are then expressed as the following
\begin{align}
    \mathcal{L} _2 = &G_2(\phi,X) \,,\\
    \mathcal{L} _3 = &-G_3(\phi,X) \square \phi \,,\\
    \mathcal{L} _4 = &G_4(\phi,X) R + G_{4,X} \left[ (\square \phi)^2 + \nabla _{\mu}\nabla _\nu \phi \nabla ^\mu \nabla ^\nu \phi\right] \,,\\
    \mathcal{L} _5 = &G_5(\phi,X) G_{\mu\nu}\nabla ^\mu \nabla ^\nu \phi - \nonumber \\
    &-\frac{1}{6} G_{5,X} \left[(\square \phi)^3 -3 \square \phi \nabla _\mu \nabla _\nu \phi \nabla ^\mu \nabla ^\nu \phi + 2 \nabla ^\mu \nabla _\alpha \phi \nabla ^\alpha \nabla _\beta \phi \nabla ^\beta \nabla _\mu \phi\right]\,,
\end{align}
where $G_i(\phi, X)$ are arbitrary functions of the scalar field and its kinetic term, $X = - \frac{1}{2}\nabla ^\mu\phi \nabla _\mu \phi$, $\square \phi = \nabla ^\mu \nabla _\mu \phi$ is the d'Alembertian operator and $G_{\mu\nu}$ is the Einstein tensor. Since the $G_i$ functions are arbitrary, it turns out that many modified theories of gravity are subclasses of Horndeski; Brans-Dicke theory can be mapped to \eqref{eq:Horndeski-action} if we choose $G_2 = 2\omega X / \phi ,\,G_3 = 0 = G_5$ and $G_4 = \phi$; while $f(R)$ gravity can be retrieved for $G_2 = f(\phi) - \phi f'(\phi), G_3 = 0 = G_5$ and $G_4 = f'(\phi).$

As already mentioned, part of the Horndeski action was severely constrained after GW170817, and specifically, $G_5(\phi,X) $ was forced to be equal to a constant, while $G_4(\phi,X)$ should only be a function of $\phi$ \cite{Ezquiaga:2017ekz}. Based on that, we decided to focus our attention on three models that still pass the GW constraint test, and these are the following,
\begin{itemize}
    \item Quintessence \cite{Tsamis:1997rk}
\begin{equation}\label{eq:quintessence}
    G_2 = X - V(\phi), G_3 = C\,, G_4 = 1/2 \,\, \text{and} \,\,G_5 = 0\,,
\end{equation}
    \item Designer Horndeksi \cite{Arjona:2019rfn}
\begin{equation}\label{eq:designer-Horndeski}
    G_2 = K(X)\,, G_3 = G(X), G_4 = 1/2\,\, \text{and}\,\,G_5 = 0\,,
\end{equation}
    \item Tailoring Horndeski \cite{Bernardo:2019vln}
\begin{equation}\label{eq:tailoring-Horndeski}
    G_2 = X - 2\Lambda \,, G_3 = G(X), G_4 = 1/2\,\,\text{and}\,\,G_5 = 0\,.
\end{equation}
\end{itemize}
$V(\phi)$ is the quintessence potential, $C$ is a constant, $K(X)$ and $G(X)$ are arbitrary functions of the kinetic term of the scalar field and $\Lambda$ is the cosmological constant. In the matter part of the action \eqref{eq:Horndeski-action} we consider a perfect fluid with energy density $\rho$ and pressure $P$.

\section{Artificial Neural Networks}\label{sec:ann_intro} 

In this section, we describe briefly the method by which ANN architectures \cite{2015arXiv151107289C} are adopted together with our technique to use them as a vehicle for forming the Hubble diagram and its derivatives in order to reconstruct particular classes of scalar-tensor theories. An ANN is structured with input and output layers being the input and output parameter interfaces while a series of hidden layers are optimized to best mimic the real data processes that result in input data producing output data values. Each layer is composed of neurons that are connected to other neurons in other layers.

\tikzset{%
  every neuron/.style={
    circle,
    fill=green!70,
    minimum size=32pt, inner sep=0pt
  },
  mid neuron/.style={
    circle,
    fill=blue!70,
    minimum size=32pt, inner sep=0pt
  },
  last neuron/.style={
    circle,
    fill=red!70,
    minimum size=32pt, inner sep=0pt
  },
  neuron missing/.style={
    draw=none,
    fill=none,
    scale=4,
    text height=0.333cm,
    execute at begin node=\color{black}$\vdots$
  },
}
\begin{figure}[t!]
    \centering
    \begin{tikzpicture}[shorten >=1pt,->,draw=black!50, node distance=\layersep]
    \tikzstyle{annot} = [text width=5em, text centered]

\foreach \m/\l [count=\y] in {1}
  \node [every neuron/.try, neuron \m/.try] (input-\m) at (0,-1.5*\y) {};

\foreach \m [count=\y] in {1,2,3,missing,4}
  \node [mid neuron/.try, neuron \m/.try ] (hidden-\m) at (5,2-\y*1.5) {};

\foreach \m [count=\y] in {1,2}
  \node [last neuron/.try, neuron \m/.try ] (output-\m) at (10,1.25-2*\y) {};

\foreach \name / \y in {1}
    \path[yshift=-.1cm] node[above] (input+\name) at (0,-1.6\name) {\large$z$};

\foreach \l [count=\i] in {1,2,3,k}
  \node[below] at (hidden-\i) {\large$\mathfrak{n}_\l$};

\foreach \name / \y in {{\large $\Upsilon(z)$} / 1, {\large$\sigma_\Upsilon^{}(z)$} / 2}
        \path[yshift=-.1cm] node[above, right of=H-3] 
        (output-\y) at (7.5,1.35-2*\y) {\name};

\foreach \i in {1}
  \foreach \j in {1,2,3,...,4}
    \draw [->] (input-\i) -- (hidden-\j);

\foreach \i in {1,2,3,...,4}
  \foreach \j in {1,2}
    \draw [->] (hidden-\i) -- (output-\j);

\foreach \l [count=\x from 0] in {\large Input, \large Hidden, \large Output}
  \node [align=center, above] at (\x*5,2) {\l \\ \large layer};

\end{tikzpicture}
    \caption{The adopted ANN architecture is shown, where the input is the redshift of a cosmological parameter $\Upsilon(z)$, and the outputs are the corresponding value and error of $\Upsilon(z)$.}
\label{fig:ANN_structure}
\end{figure}
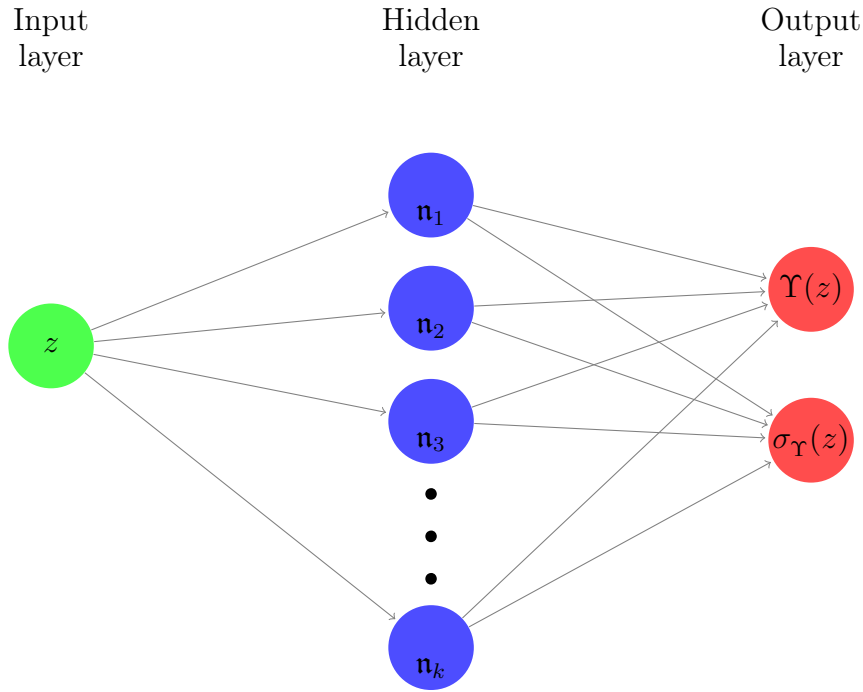

In our case, the input layer will accept redshift values while the output layer gives the Hubble parameter for that redshift and its uncertainty. In this way, an input signal, or redshift value, traverses the whole network to produce these outputs. To illustrate this architecture, we show a one hidden layer ANN for a generic cosmological parameter $\Upsilon(z)$ and its corresponding uncertainty $\sigma_\Upsilon^{}(z)$ in Fig. \ref{fig:ANN_structure}, where the neurons are denoted by $\mathfrak{n}_k$. Here, the ANN is structured so that a linear transformation (composed of linear weights and biases) is applied for each of the different layers.

Neurons are composed of an activation function that, across the larger number of neurons, can be used to model the complex relationships that feature in the data. In our work, we used the Exponential Linear Unit (ELU) \cite{2015arXiv151107289C} as the activation function, specified by
\begin{equation}
    f(x) = 
  \begin{cases} 
   {x} & \text{if } x>0 \\
   {\alpha(e^x-1)} & \text{if } x \leq 0
  \end{cases}\,,
\end{equation}
where $\alpha$ is a positive hyperparameter that controls the value to which an ELU saturates for negative net inputs, which we set to unity. Besides being continuous and differentiable, the function does not act on positive inputs while negative inputs tend to be closer to unity for more negative input values.

The linear transformations and activation function produce a huge number of so-called hyperparameters which are nonphysical and can be optimized using training data so that the larger system mimics some physical process as closely as possible. The process by which training takes place involves these hyperparameters being optimized by comparing the predicted result $\hat{\Upsilon}$ with the ground truth $\Upsilon$ (training data) so that their difference is minimized. This minimization is called the loss function. This function is minimized by fitting methods such as gradient descent which fixes the hyperparameters. In our work, we adopt Adam's algorithm \cite{2014arXiv1412.6980K} as our optimizer which is a modification of the gradient descent method that has been shown to accelerate convergence. 

The direct absolute difference between the predicted ($\hat{\Upsilon}$) and training ($\Upsilon$) outputs summed for every redshift is called the L1 loss function. This is akin to the log-likelihood function for uncorrelated data in a Markov Chain Monte Carlo (MCMC) analysis and is a very popular choice for ANN architectures. There are other choices such as the mean squared error (MSE) loss function which minimizes the square difference between $\hat{\Upsilon}$ and $\Upsilon$, and the smooth L1 (SL1) loss function which uses a squared term if the absolute error falls below unity and absolute term otherwise. Thus, it is at the level of the loss function that complexities in the data are inserted into the ANN through optimization in the training of the numerous hyperparameters that make up the system. In our work, we do this by assuming a loss function that is more akin to correlated data when considering log-likelihood functions for MCMC analyses. To that end, we incorporate the covariance matrix $\mathrm{C}$ of a data set by taking a loss function, given as
\begin{equation}
    {\rm L_{\chi^2}} = \sum_{i,j} \left[H_{\rm obs}(z_i) - H_{\rm pred}(z_i)\right]^\text{T} \mathrm{C}_{ij}^{-1} \left[H_{\rm obs}(z_j) - H_{\rm pred}(z_j)\right] \,, \label{eq:chi2_loss}
\end{equation}
where $\mathrm{C}_{ij}$ is the total noise covariance matrix of the data, which includes the statistical noise and systematics.

Additionally, we wish to perform reconstruction on scalar-tensor classes of theories which also require the $H'(z)$ parameter. One direct approach is to consider the numerical derivative of the Hubble diagram. However, the uncertainties using this approach are unreasonably large. Taking the same approach as Ref.~\cite{Mukherjee:2022yyq}, we can consider a Monte Carlo approach by which simulated points at every redshift are considered using numerical differentiation. In turn, the mean and uncertainties can be obtained directly. Using this technique, we are able to reconstruct not only $H(z)$ but also its derivative $H'(z)$.

ANNs that feature at least one hidden layer can approximate any continuous function for a finite number of neurons, provided the activation function is continuous and differentiable \cite{HORNIK1990551}, which means that ANNs are applicable to the setting of cosmological data sets. In this work, we utilize the code for reconstructing functions from data called Reconstruct Functions with ANN (\texttt{ReFANN}\footnote{\url{https://github.com/Guo-Jian-Wang/refann}}) \cite{Wang:2019vxv} which is based on \texttt{PyTorch}\footnote{\url{https://pytorch.org/docs/master/index.html}}. The code was run on GPUs to speed up the computational time, as well as making use of batch normalization \cite{2015arXiv150203167I} prior to every layer which further accelerates the convergence.

\section{Reconstruction of the Hubble Evolution using ANNs} \label{sec:ANN_results}

We now employ ANNs to reconstruct the Hubble diagram, considering three sources of $H(z)$ data. These include the cosmic chronometers (CC), type Ia supernovae (SN), and baryonic acoustic oscillation (BAO) measurements. Furthermore, keeping in mind the rising $H_0$ tension, we consider the most precise Cepheid calibration result of $H_0 =  73.3 \pm 1.04$ km Mpc$^{-1}$ s$^{-1}$ \cite{Riess:2021jrx} by the SH0ES team (hereafter referred to as R21), recently inferred $H_0 = 69.7 \pm 1.9$ km Mpc$^{-1}$ s$^{-1}$ \cite{Freedman:2021ahq} via the Tip of the Red Giant Branch (TRGB) calibration technique (hereafter referred to as TRGB) and the most precise early-time determination of $H_0 = 67.4 \pm 0.5$ km Mpc$^{-1}$ s$^{-1}$ \cite{Aghanim:2018eyx} inferred from the Cosmic Microwave Background (CMB) sky by the Planck 2018 survey (hereafter referred to as P18).  In our analysis, we assume Gaussian prior distributions with the mean and variances corresponding to the central and 1$\sigma$ reported values of each prior above.

The latest 32 CC $H(z)$ measurements \cite{Stern:2009ep,Moresco:2012jh, Moresco:2016mzx, Borghi:2021rft, Ratsimbazafy:2017vga, Moresco:2015cya, Zhang:2012mp}, covering the redshift range up to $z \sim 2$, do not assume any particular cosmological model but depend on the differential ages technique between galaxies, where we consider the full covariance matrix including the systematic and calibration errors as reported by Moresco\cite{Moresco:2020fbm}. Furthermore, we take into account the Hubble distance $\frac{d_H(z)}{r_d}$ and the transverse comoving distance $\frac{d_M(z)}{r_d}$ BAO measurements \cite{BOSS:2016wmc,Bautista:2020ahg,Gil-Marin:2020bct,Tamone:2020qrl,deMattia:2020fkb,Neveux:2020voa,Hou:2020rse,deSainteAgathe:2019voe,Blomqvist:2019rah} from different galaxy surveys like Sloan Digital Sky Survey (SDSS), the Baryon Oscillation Spectroscopic Survey (BOSS) and the extended Baryon Oscillation Spectroscopic Survey (eBOSS), such that
\begin{eqnarray}
    {d_H(z)} &=& {c}/{H(z)}\,, \\
    {d_M(z)} &=& {(1+z)}{d_A(z)}\,, 
\end{eqnarray} 
where $d_A(z)$ is the angular-diameter distance. For the SN data, we employ the compressed Pantheon compilation together with the CANDELS and CLASH Multi-cycle Treasury (MCT) measurements \cite{Riess:2017lxs}. When incorporating the compressed Pantheon + MCT $E(z)$ data set we compute $H(z) = H_0 E(z)$ using the three different $H_0$ priors and then feeding the resulting $H(z)$ along with the corresponding covariance matrix for training the network.

\begin{figure}
\centering
\includegraphics[width=\textwidth]{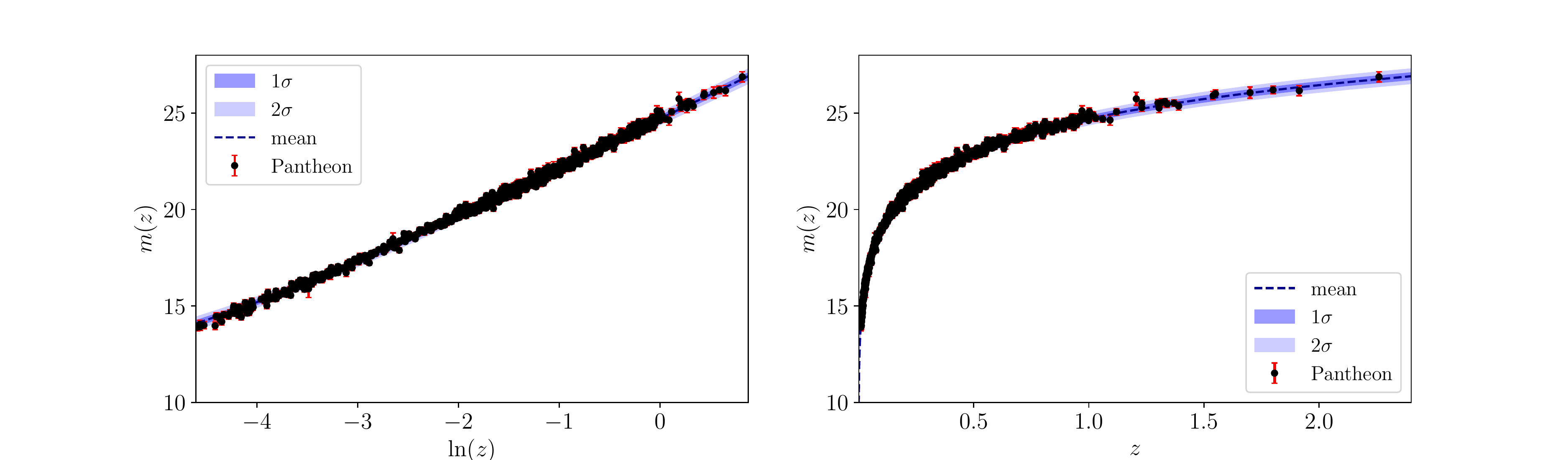}
\caption{ANN reconstruction of the Pantheon supernova apparent magnitudes $m(z)$ as a function of the $\ln(z)$ (left) and $z$ (right).} \label{m_recon}
\end{figure}

\begin{figure}
\centering
\includegraphics[width=0.45\textwidth]{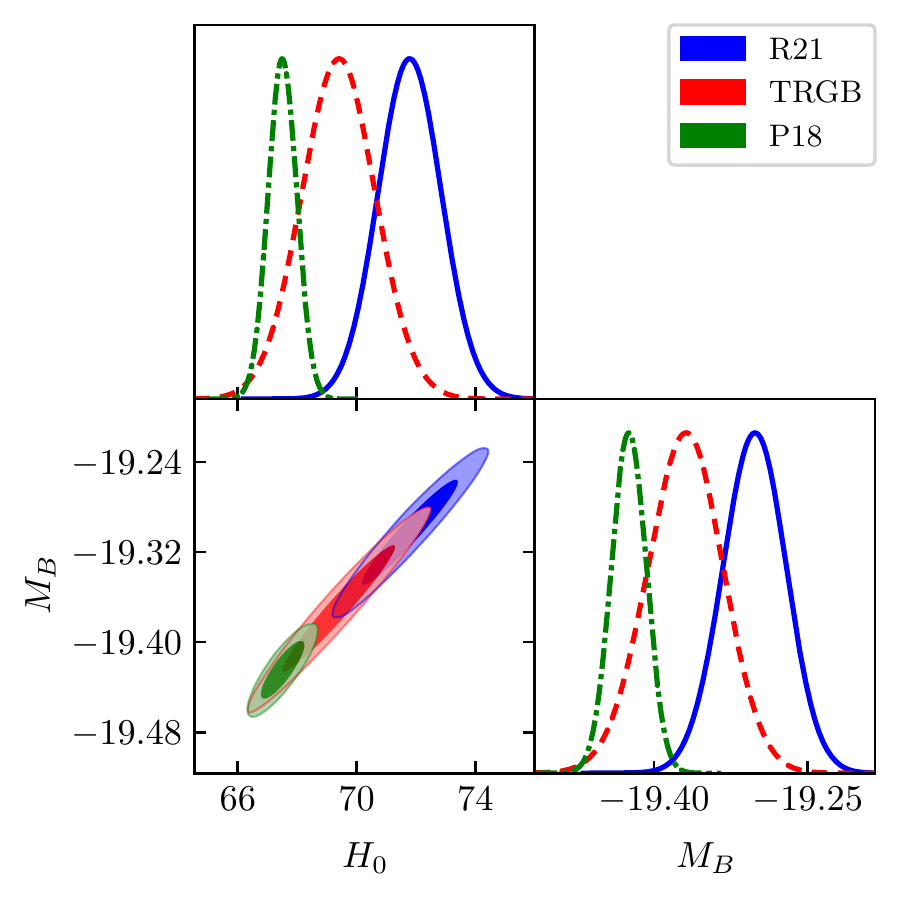} ~~~~ \includegraphics[width=0.45\textwidth]{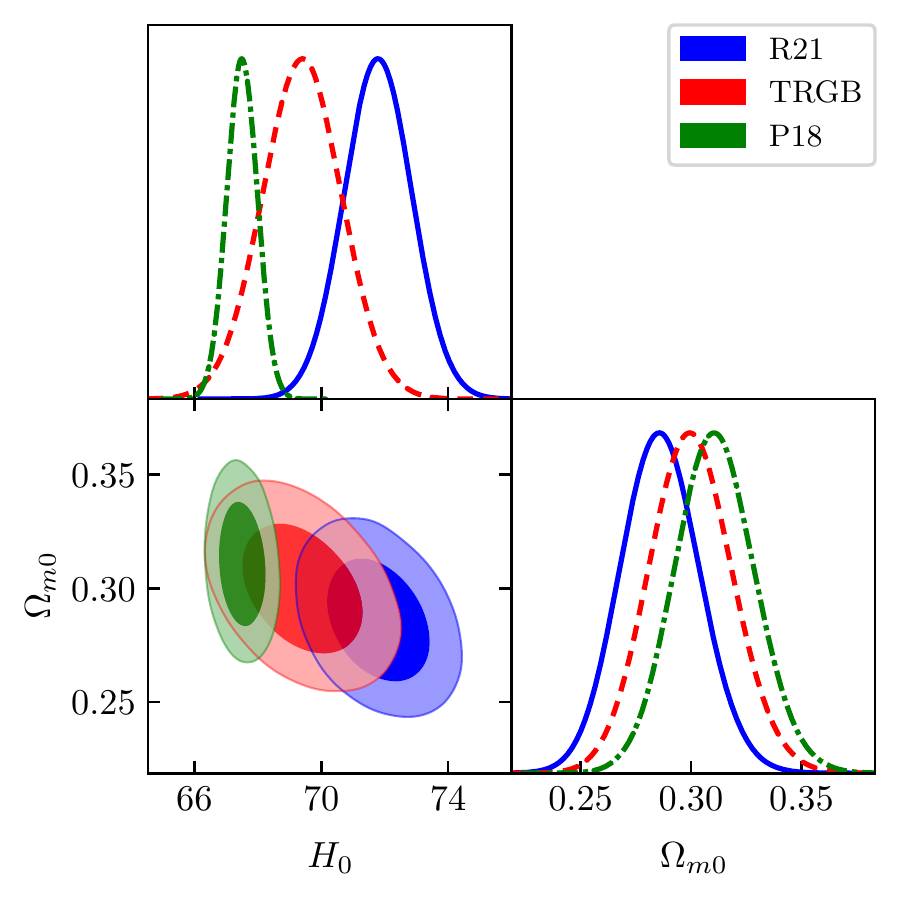} 
\caption{Marginalized posteriors for the calibrated values of supernovae apparent magnitude $M_B$ in the Pantheon compilation (in the left panel) along with the matter density parameter at present epoch $\Omega_{m0}$ (in the right panel), considering the R21, TRGB, and P18 $H_0$ priors (in units of km Mpc$^{-1}$ s$^{-1}$), respectively. The constraints obtained are $M_B$ = $-19.302 \pm 0.031$, $-19.369 \pm 0.037$ and $-19.425\pm 0.017$; and $\Omega_{m0} = 0.286\pm 0.018, ~ 0.301 \pm 0.019~ \text{and } 0.311 \pm 0.017$, corresponding to the R21, TRGB and P18 $H_0$ priors.} \label{MB_calib}
\end{figure}

A neural network reconstruction of the apparent magnitudes $m(z)$ from the full Pantheon \cite{Pan-STARRS1:2017jku} compilation is then carried out in $\ln(z)$. This is done to scale down the drastic variance in the density of data with redshift, and the result is shown in Fig. \ref{m_recon}. To undertake this reconstruction, we incorporate the total covariance matrix of a Pantheon data set and train the network by minimizing the loss function, defined in Eq. \eqref{eq:chi2_loss}. Next, we make predictions by feeding the sequence of redshifts to the input layer to obtain the $m(z)$ vs $z$ reconstruction profile. We repeat this exercise to obtain 20 realizations of the reconstructed $m(z)$ function for different initialization of the weights and biases associated with this network model. From these reconstructed $m(z)$ samples, we obtain the best-fit values of reconstructed $m(z)$ along with the associated confidence levels using a Monte Carlo routine. 

To minimize the model dependence associated with the BAO measurements we follow a similar prescription in \cite{Bernardo:2021qhu}, instead of assuming a fiducial radius of the comoving sound horizon $r_d = 147.78$ Mpc \cite{Aghanim:2018eyx}. We calculate the ratio $d_M/d_H(z)$ from the BAO data and complement this with the ANN reconstructed $m(z)$ from the full Pantheon sample, as
\begin{eqnarray}
    d_A(z) &=& d_L(z) (1+z)^{-2}, \\
    m(z) &=& 5 \log_{10} d_L(z) + 25 + M_B,
\end{eqnarray} 
where $d_L(z)$ is the luminosity distance and $M_B$ is the absolute magnitude of supernovae, assuming spatial isotropy and the cosmic distance-duality relation holds true. We obtain the marginalized constraints on $M_B$ and the matter density parameter $\Omega_{m0}$ assuming the vanilla $\Lambda$CDM model, considering a uniform prior $M_B \in [-35, -5]$, $\Omega_{m0} \in [0.01,0.9]$ via an MCMC analysis using {\texttt{emcee}}\footnote{\url{https://github.com/dfm/emcee}} \cite{Foreman-Mackey:2012any} python library. The calibrated constraints obtained are $M_B$ = $-19.302 \pm 0.031$, $-19.369 \pm 0.037$ and $-19.425\pm 0.017$ corresponding to the R21, TRGB and P18 $H_0$ priors, respectively, are shown in Fig. \ref{MB_calib} using {\texttt{GetDist}}\footnote{\url{https://github.com/cmbant/getdist}} \cite{Lewis:2019xzd}. Finally, we can evaluate the Hubble parameter measurements from the BAO data, as
\begin{equation}
    H(z) = \frac{d_M}{d_H} \cdot 10^{\frac{25 + M_B - m}{5}}\,.
\end{equation}

\begin{figure}[!t]
\centering
\includegraphics[width=0.485\textwidth]{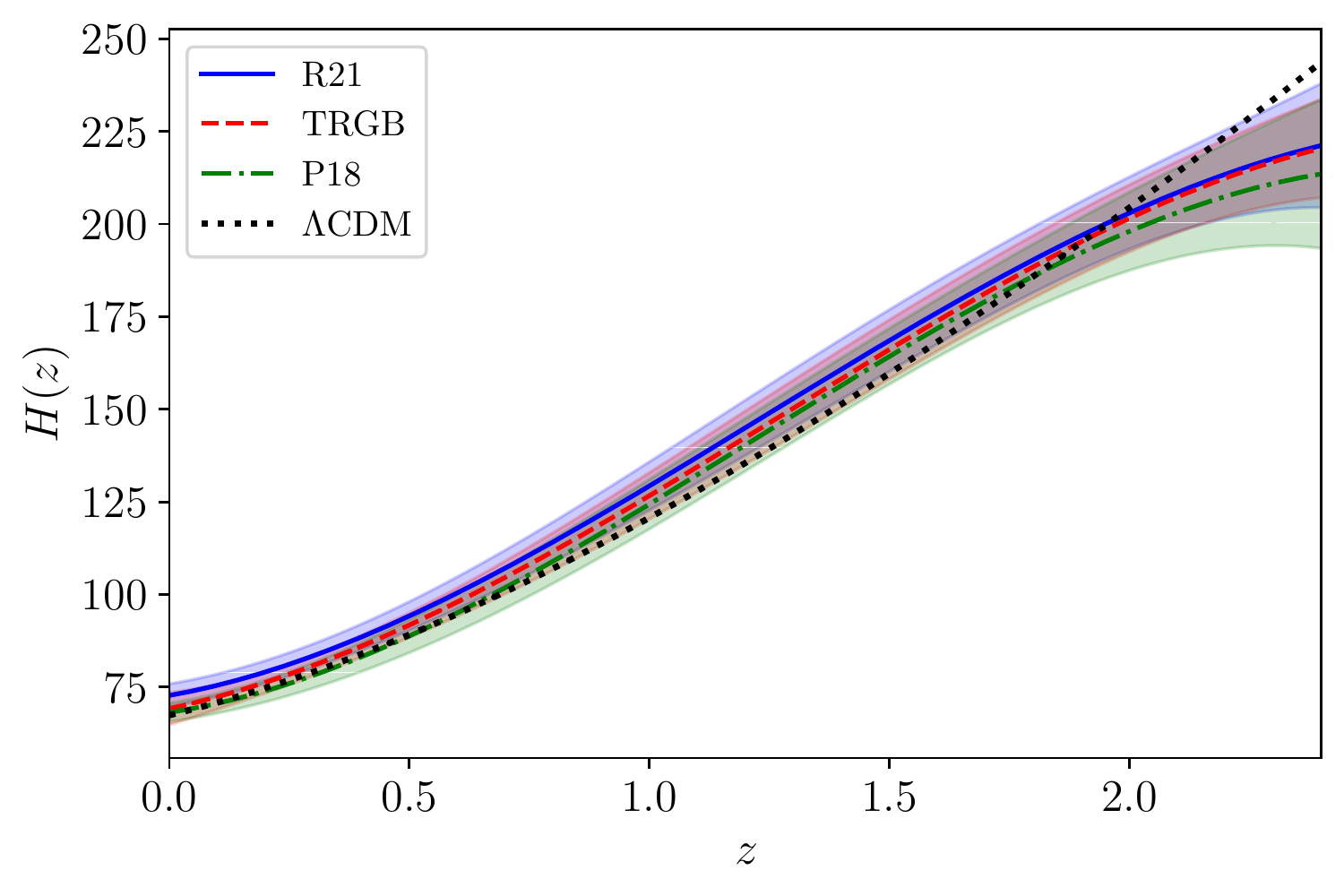} \hfill \includegraphics[width=0.485\textwidth]{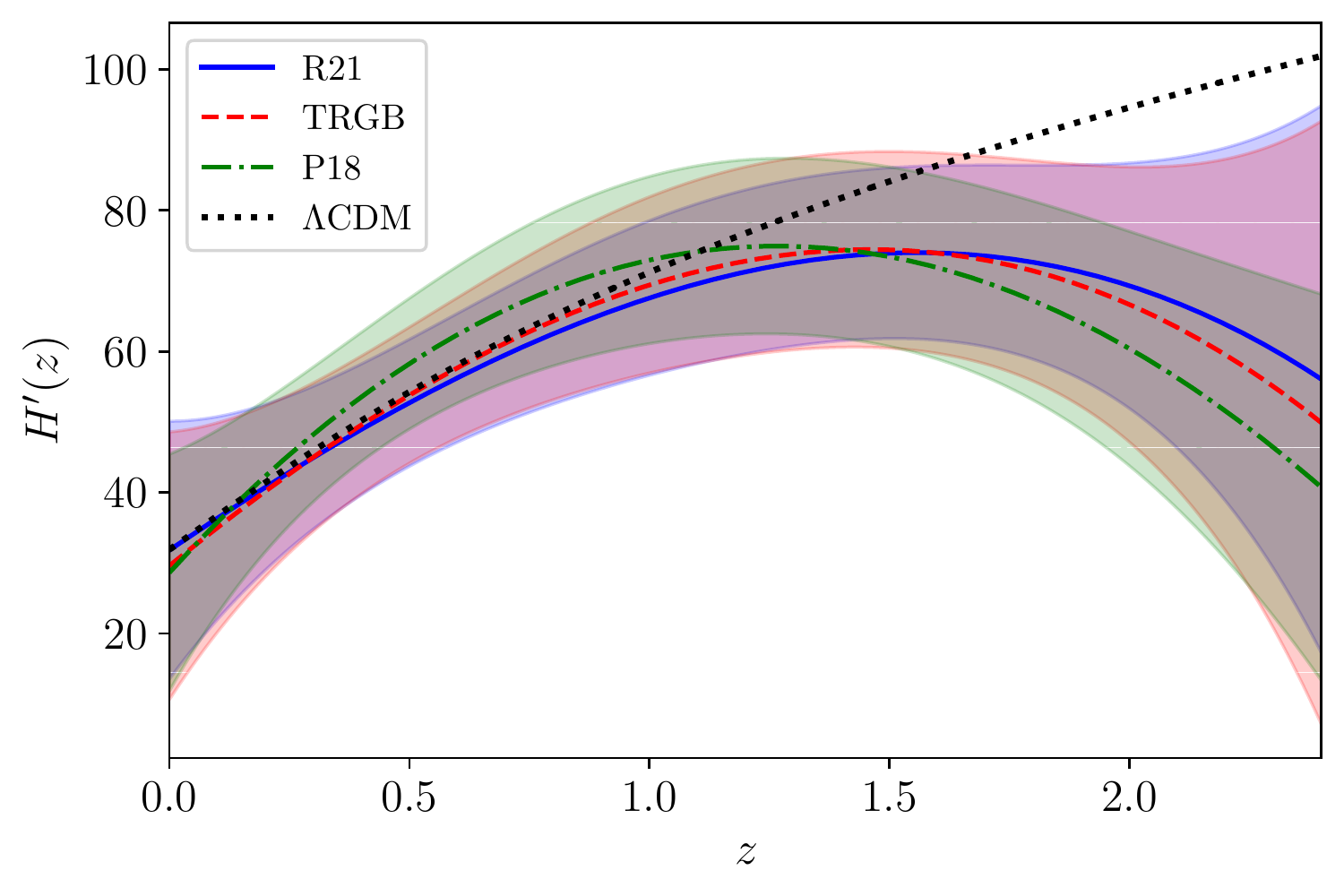}
\caption{Plots for the reconstructed $H(z)$ and $H'(z)$ using neural networks, with the Hubble data  considering R21, TRGB, and P18 $H_0$ priors. The shaded regions with `$-$', `$\vert$' and `$\times$' hatches represent the 2$\sigma$ confidence levels for the R21, TRGB, and P18 $H_0$ priors respectively.} \label{H_recon}
\end{figure}

After preparation of the Hubble data, we utilize the ANN method to reconstruct the Hubble parameter $H(z)$. Before training the network on real data, we structure the ANN for setting up the optimal network configuration, i.e. determining the optimal number of neurons and layers, for which we make use of a mock simulated Hubble dataset, generated assuming a fiducial $\Lambda$CDM model. Next, for the given sample of Hubble data, we train a network model (as described in Sec. \ref{sec:ann_intro}) to learn to mimic the complex relationships between $z$, $H(z)$ and $\sigma_H(z)$. With this trained model, any arbitrary number of $H(z)$ samples can be reconstructed by feeding a sequence of redshifts to this network model.

We repeat this exercise and obtain 1000 realizations of the Hubble function for different initialization of the weights and biases associated with the network model. From these 1000 reconstructed $H(z)$ samples, we obtain the best-fit values of reconstructed $H(z)$ along with the associated confidence levels using a Monte Carlo routine. Moreover, we also undertake the simultaneous reconstruction of $H'(z)$, where this prime denotes derivative with respect to the redshift $z$, via an MC routine on 1000 realizations of the Hubble diagram. This compounding effect of MC with ANNs is undertaken following the methodology described in Ref. \cite{Mukherjee:2022yyq}. 

The reconstructed $H(z)$ and $H'(z)$ functions are shown in Fig. \ref{H_recon}. In the next section, the reconstructed $H(z)$ and its first derivative $H'(z)$ will be used to predict the Horndeski Lagrangian potentials. It is worth mentioning that in this work, we have not delved into the effect of correlations between the reconstructed functions, $H(z)$ and $H'(z)$, for a simplified analysis.

\section{Data-driven Scalar-Tensor Models \label{sec:model_results}}

The structure of the spacetime is considered to be a homogeneous and isotropic, spatially flat Friedmann–Lema\^{i}tre–Robertson–Walker (FLRW) Universe of the form
\begin{equation}\label{eq:FLRW-metric}
    \dd s^2 = - \dd t^2 + a^2(t) (\dd x ^2 + \dd y ^2 + \dd z ^2 )\,,
\end{equation}
where $a(t)$ is the scale factor, on which the Hubble parameter $H(t) = \dot{a}(t)/a(t)$ is built.

\subsection{Quintessence} \label{sec:quint_res}

\begin{figure}[!t]
\centering
\includegraphics[width=0.485\textwidth]{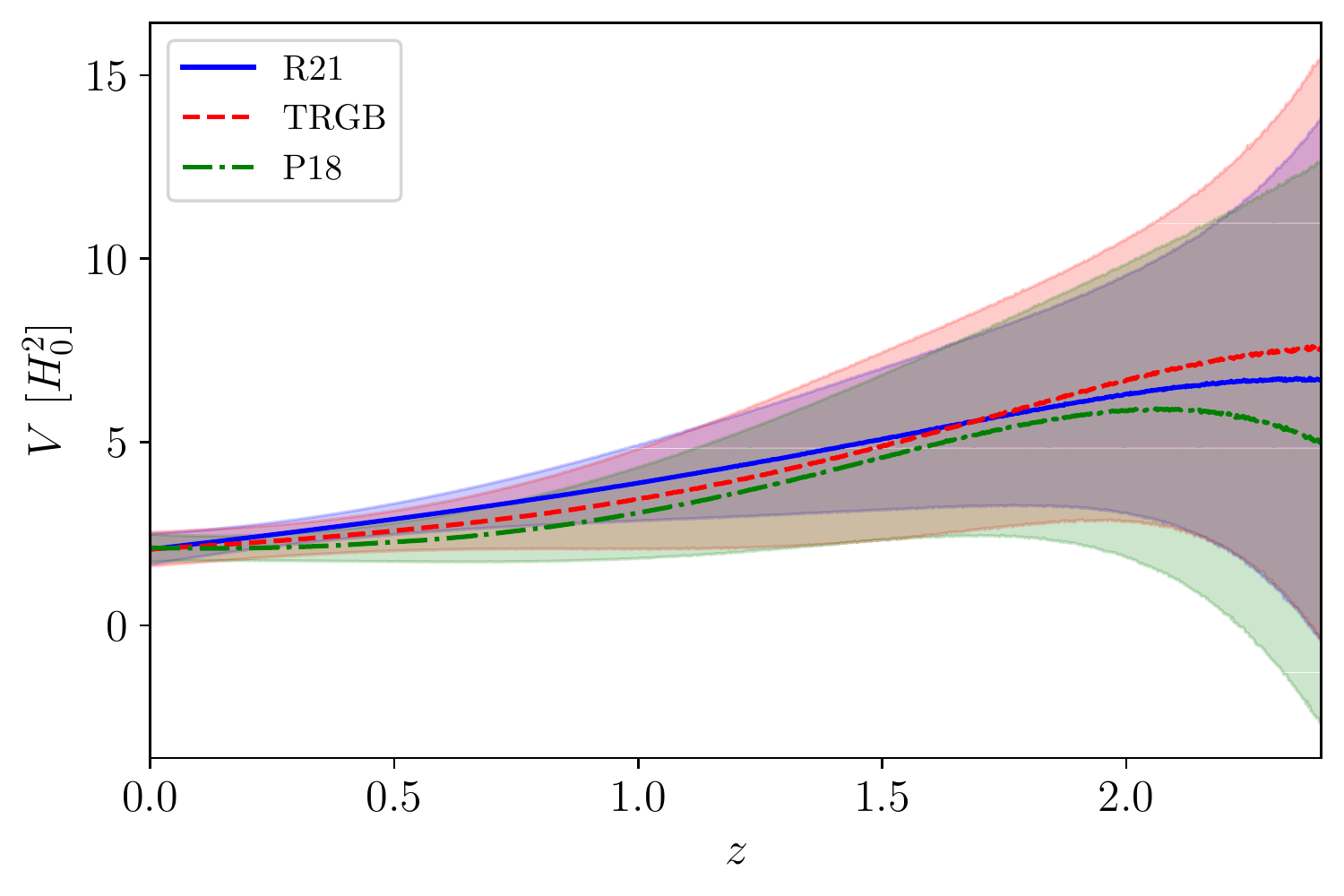} \hfill \includegraphics[width=0.485\textwidth]{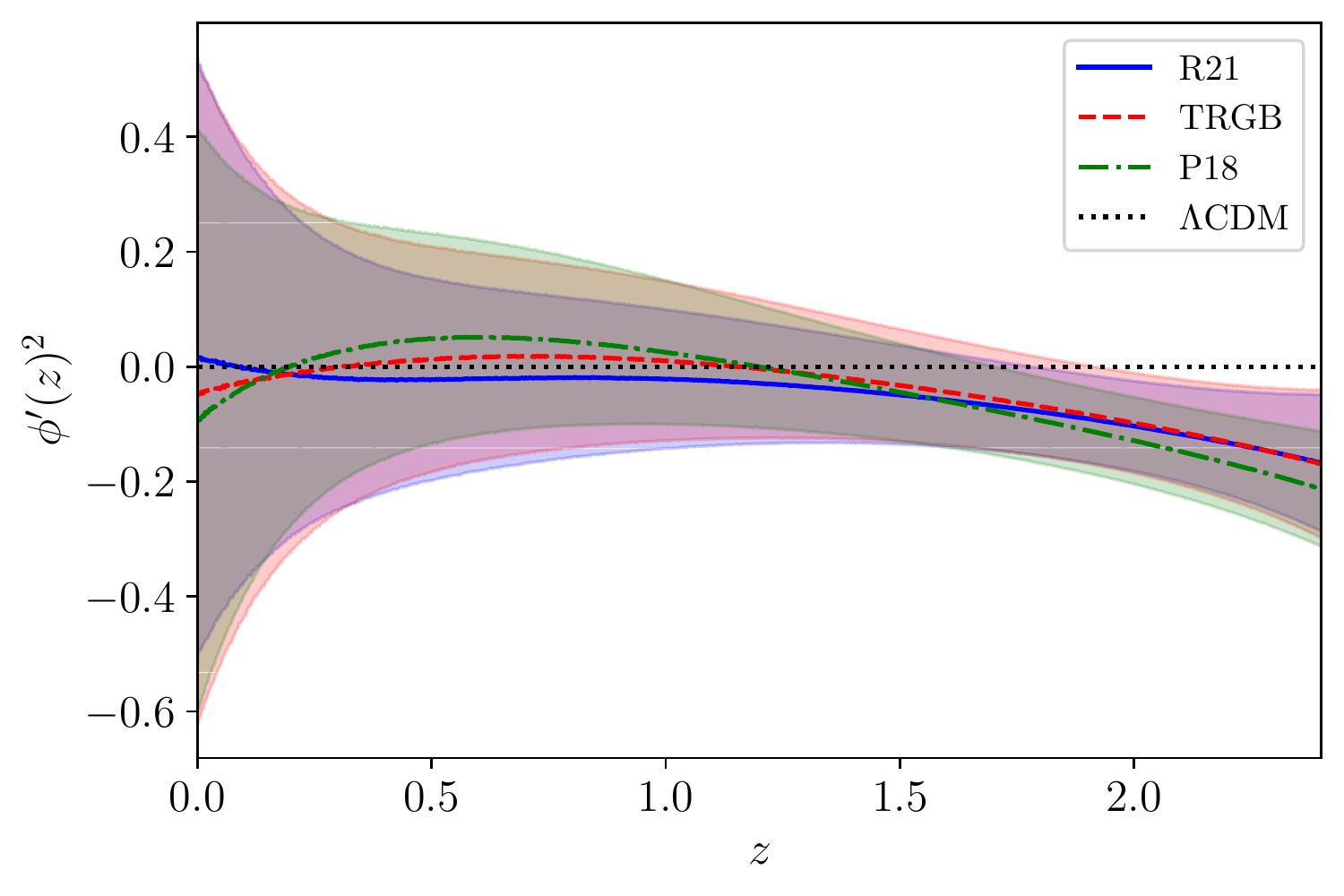}
\caption{Plots for the reconstructed $V(z)$ (in left panel) and $\phi^{\prime}{}^2(z)$ using neural networks considering R21, TRGB, and P18 $H_0$ priors. The shaded regions with `$-$', `$\vert$' and `$\times$' hatches represent the 2$\sigma$ confidence levels for the R21, TRGB, and P18 $H_0$ priors respectively.} \label{fig:quint}
\end{figure}

Varying the action in Eq.~\eqref{eq:Horndeski-action} with $G_i$ being the ones in \eqref{eq:quintessence} and the metric \eqref{eq:FLRW-metric}, we get the Friedmann equations as
\begin{gather}
    3 H^2 = \rho +\frac{\dot{\phi}^2}{2} + V(\phi)\,,\\
    2\dot{H} + 3 H^2 = - P - \frac{\dot{\phi}^2}{2} + V(\phi)\,,
\end{gather}
while the Klein-Gordon equation reads
\begin{eqnarray}
    \ddot{\phi} + 3 H \dot{\phi} + V'(\phi) = 0\,,
\end{eqnarray}
where $V(\phi)$ represents the scalar field potential. Solving for the quintessence potential and the kinetic term of the scalar field, we get
\begin{gather}
    V(\phi) = \dot{H} + 3 H^2 - \frac{\rho - P}{2}\,, \label{eq:V_quint} \\
    \dot{\phi}^2 = -2 \dot{H} - (\rho -P)\, .  \label{eq:phi_quint}
\end{gather}
Notice that given information on the matter fields and the Hubble evolution, we can determine both the potential and the kinetic term of the quintessence field. The matter sources are considered to be non-relativistic, i.e. $P = 0$, and the value of the current matter density parameter $\Omega _{m0} = \frac{8 \pi G_{\rm N}}{3 H_0 ^2} \rho_{m} = 0.286\pm 0.018, ~ 0.301 \pm 0.019~ \text{and } 0.311 \pm 0.017$, corresponding to the R21, TRGB and P18 $H_0$ priors, respectively, shown in Fig. \ref{MB_calib}. We use three different priors in order to avoid possible bias.

We can now use the reconstructed $H(z)$ and $H'(z)$ to predict the functions, $V$ and $\phi^{\prime}{}^2$, where $\dot{\phi}(t) = -(1+z)H(z) \phi^\prime(z)$. The results are shown in Fig. \ref{fig:quint}. We have adopted a dimensionless way of expressing the potential and kinetic terms, in units of $H_0$, to alleviate the tensions existing at $z = 0$ for the different $H_0$ priors. Therefore, we find that the mean reconstructed curves have a significant overlap at the 2$\sigma$ confidence level, irrespective of the $H_0$ prior considered. 

\begin{figure}[!t]
\centering
\includegraphics[width=0.485\textwidth]{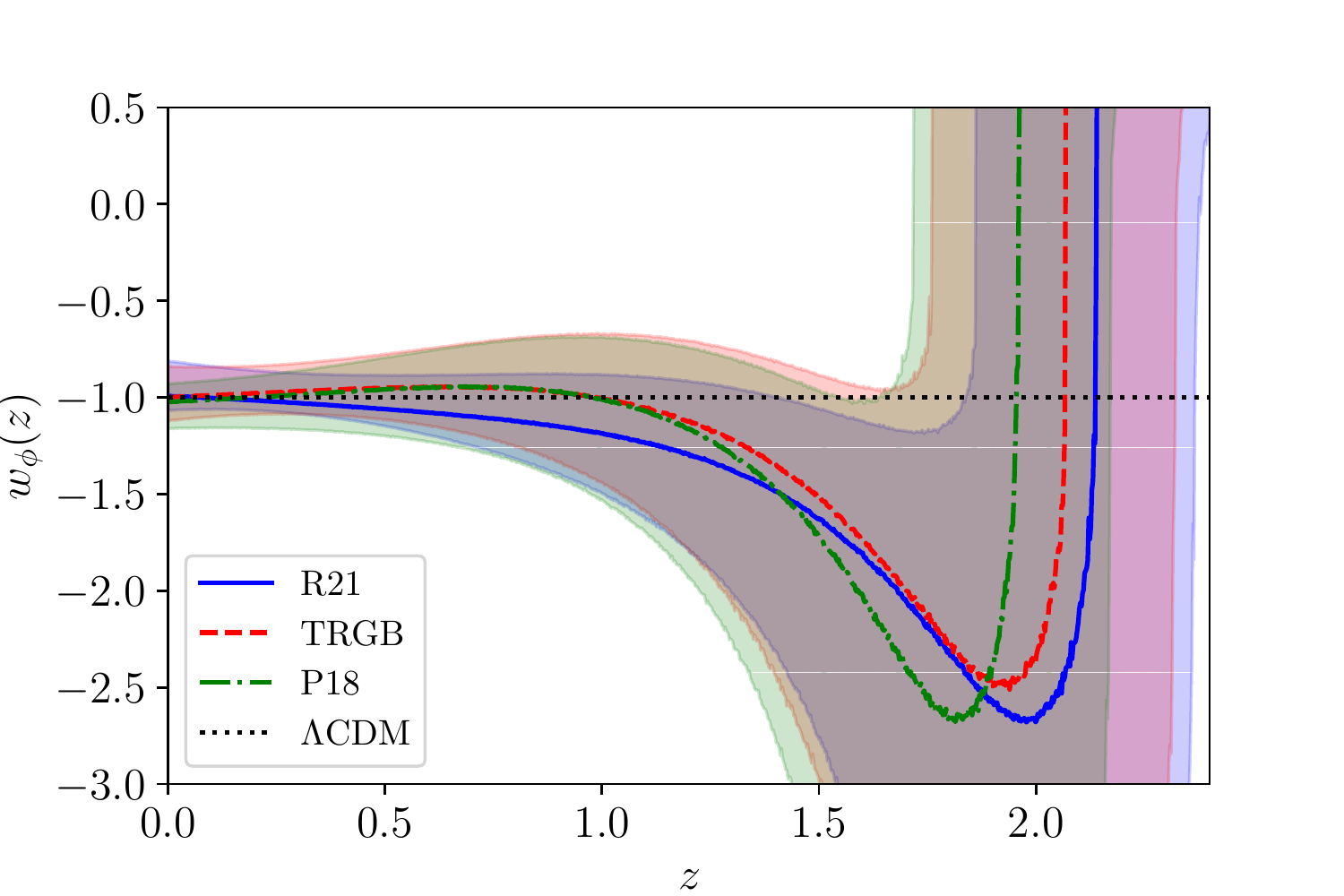} \hfill \includegraphics[width=0.485\textwidth]{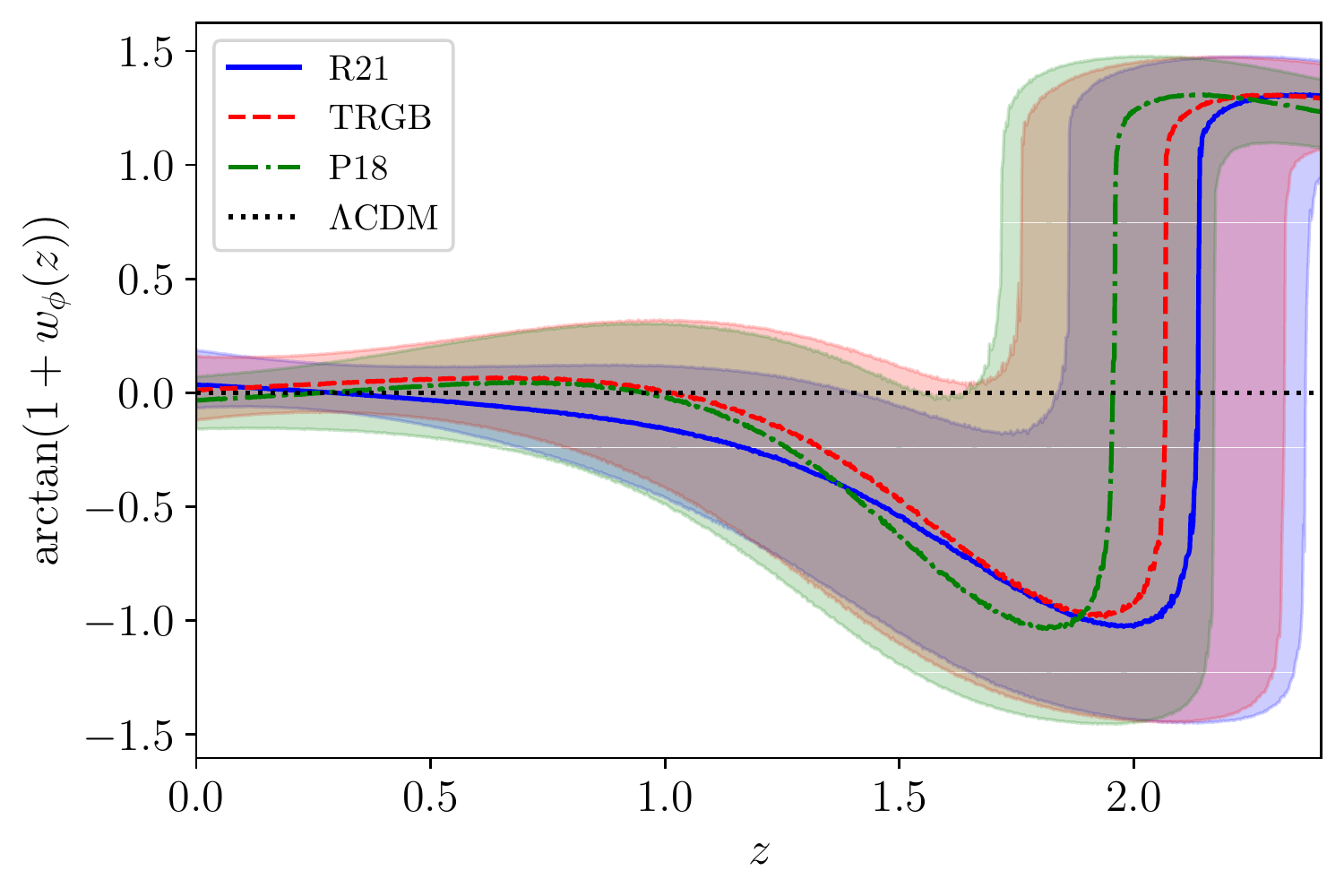}
\caption{Plots for the quintessence dark energy EoS $w_\phi(z)$ (in the left panel) and its compactified form $\arctan(1+w_\phi(z))$ (in the right panel) considering R21, TRGB, and P18 $H_0$ priors. The shaded regions with `$-$', `$\vert$' and `$\times$' hatches represent the 1$\sigma$ confidence levels for the R21, TRGB, and P18 $H_0$ priors respectively.} \label{fig:w_quint}
\end{figure}

\begin{figure}[!t]
\centering
\includegraphics[width=0.325\textwidth]{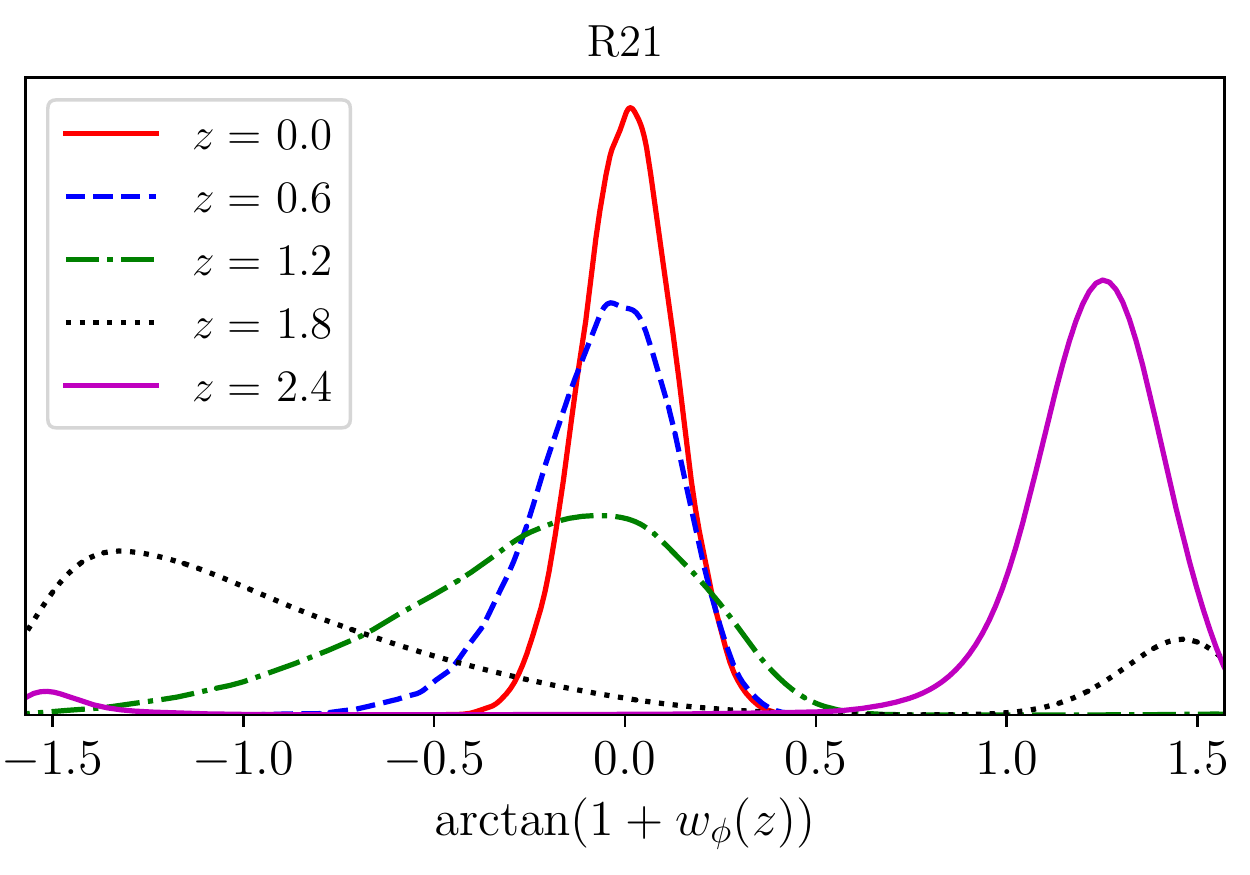} \hfill \includegraphics[width=0.325\textwidth]{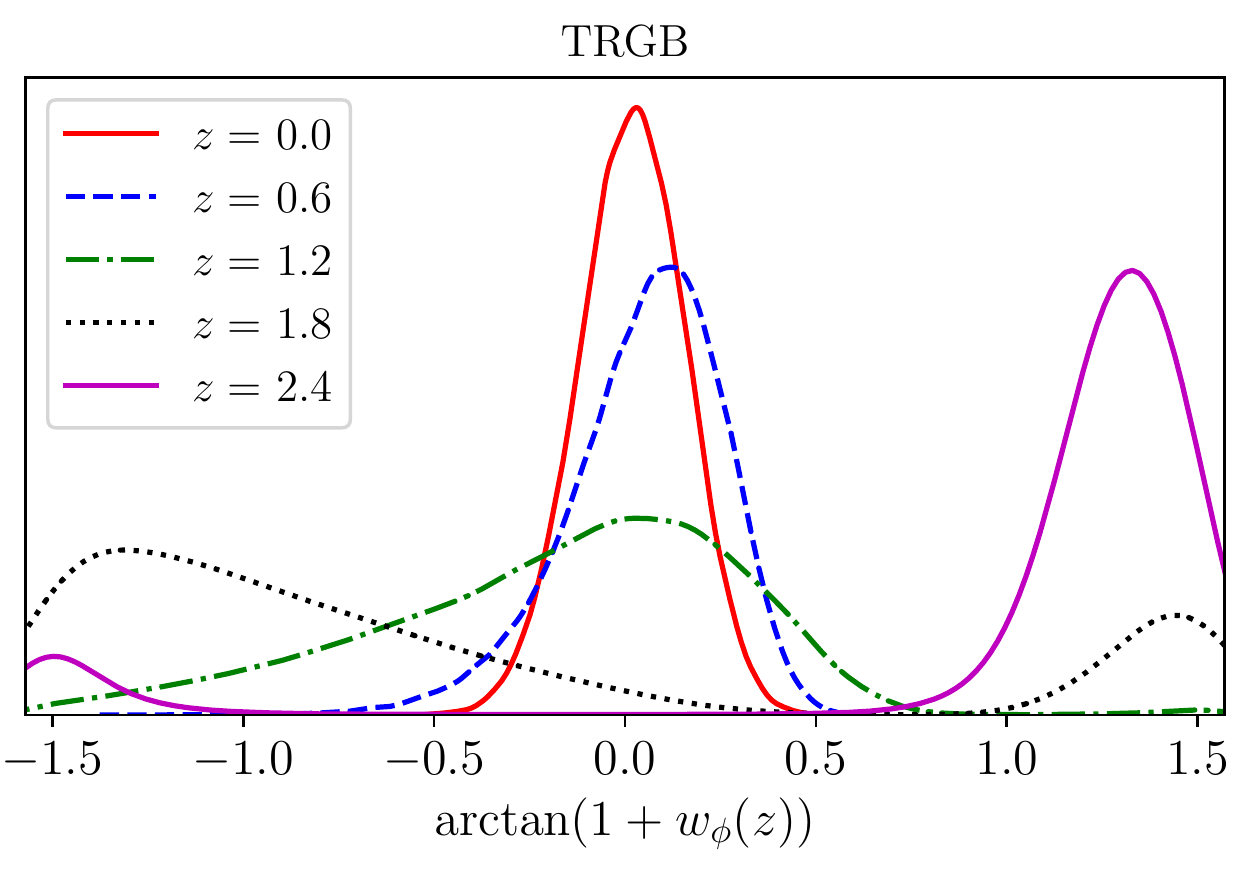} \hfill \includegraphics[width=0.325\textwidth]{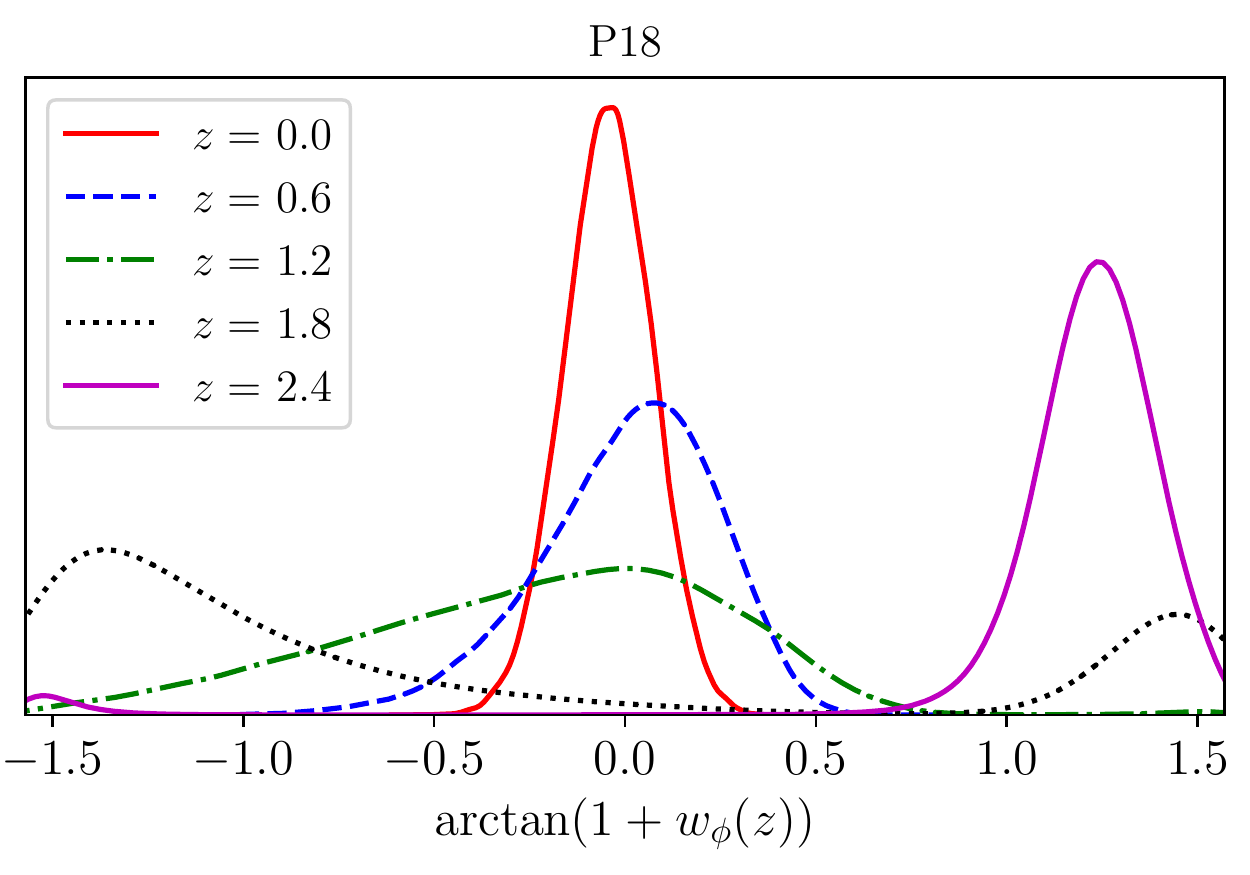}
\caption{Plots showing the posterior distribution of the compactified dark energy EoS at some sample redshifts for the R21, TRGB, and P18 $H_0$ priors, respectively.} \label{fig:atw_dist_quint}
\end{figure}

From the scalar field Eqs.~\eqref{eq:V_quint} and \eqref{eq:phi_quint}, one can arrive at the evolution of the dark energy equation of state (EoS) as
\begin{equation}
    w_\phi = \frac{\dot{\phi}^2/2 - V}{\dot{\phi}^2/2 + V}\,. \label{eq:w_de_quint}
\end{equation}
For convenience, we sample over the compactified variable, $\arctan(1+w_\phi)$, and call it the compactified dark energy EoS. Fig. \ref{fig:w_quint} shows the results for the dark energy EoS and its compactified form. We also plot the resulting posterior distributions of compactified dark energy EoS at different redshifts in Fig. \ref{fig:atw_dist_quint}. Interestingly, we see in Fig. \ref{fig:atw_dist_quint} that for low redshifts the distribution is normal. However, for higher redshifts, where the $w_\phi$ diverges, the distribution becomes non-Gaussian.

\subsection{Designer Horndeski}

The cosmological field equations for the metric and the scalar field in designer Horndeski \eqref{eq:designer-Horndeski} are
\begin{gather}
    3 H^2 = \rho - K(X) + 2 X K_X + 3 H \dot{\phi}^2 G_X\,, \label{eq:hdes_fried1} \\
    2 \dot{H} + 3 H^2 = - P - K(X) + 2 X \ddot{\phi} G_X \,, \label{eq:hdes_fried2} \\
\ddot{\phi} \left[ \dot{\phi} \left(3 H (G_{XX}\dot{\phi}^2 + 2 G_X) + K_{XX} \dot{\phi} \right) + K_X\right] + 3 \dot{\phi} \left(G_X \dot{H} \dot{\phi} + 3 G_X H^2 \dot{\phi} + H K_X \right) = 0\,, \label{eq:hdes_fried3}
\end{gather}
where the subscript in the functions $K(X)$ and $G(X)$ denotes differentiation with respective to the kinetic term $X$.

In cubic Horndeski theory (obtained by setting $G_4 = \frac{1}{2}, ~ G_5 = 0$ in the general Horndeski action \eqref{eq:Horndeski-action}), assuming shift symmetry such that the $G_i$'s only depend on
$X$), the $k$-esssence and braiding potentials can be obtained via a designer approach by assuming $H = H(X)$ to close the system of field equations. It deserves mention that, in these shift symmetric models, the scalar field equation takes the form
\begin{equation}
    \dot{J} + 3 H J = 0\,, \label{eq:hdes_sf}
\end{equation}
where $J$ is the shift current given by
\begin{equation}
    J = \dot{\phi} K_X + 3 H \dot{\phi}^2 G_X\,. \label{eq:hdes_current}
\end{equation}
Therefore, equation \eqref{eq:hdes_sf} can be rewritten as an exact solution, such that
\begin{equation}
\dot{\phi} K_X + 3 H \dot{\phi}^2 G_X = \dfrac{\mathcal{J}}{a^3} \,, \label{eq:hdes_J}
\end{equation}
where $\mathcal{J}$ is an integration constant, often referred to as a \textit{shift charge}, which describes deviation from the dynamical attractor behaviour $J = 0$. 

The designer approach recognizes \eqref{eq:hdes_fried1} and \eqref{eq:hdes_J} as two independent equations of the system, but with three unknowns, namely, $\left\lbrace H, K(X), G(X) \right\rbrace$. For our designer Horndeski models, we assume the Hubble parameter to be related to the scalar field as,
\begin{equation}
X = \frac{c_0}{H(z)^n} , 
\end{equation}
where $c_0$, expressed in units of $H_0^{n + 2}$, and $n$ are positive constants. In this case, the potentials are given by
\begin{equation}
K(X) = -3 H_0^2 \left( 1- \Omega_{m0} \right) + \dfrac{\mathcal{J} \sqrt{2X} H^2(X)}{ H_0^2 \Omega_{m0} } - \dfrac{ \mathcal{J} \sqrt{2X} \left( 1-\Omega_{m0} \right)}{ \Omega_{m0} }\,, \label{eq:K_hdes}
\end{equation}
and
\begin{equation}
G_X(X) = - \dfrac{ 2 \mathcal{J} H'(X) }{ 3 H_0^2 \Omega_{m0} }\,. \label{eq:GX_hdes}
\end{equation}
where $\mathcal{J}$ is described in units of $H_0$, and $\Omega_{m0}$ is the dark matter density parameter at the present epoch. We have considered $\Omega _{m0} = 0.286\pm 0.018, ~ 0.301 \pm 0.019~ \text{and } 0.311 \pm 0.017$, corresponding to the R21, TRGB and P18 $H_0$ priors, similar to Sec. \ref{sec:quint_res}.\\

\begin{figure}[!t]
\center
\subfigure[ ]{
\includegraphics[width = 0.475 \textwidth]{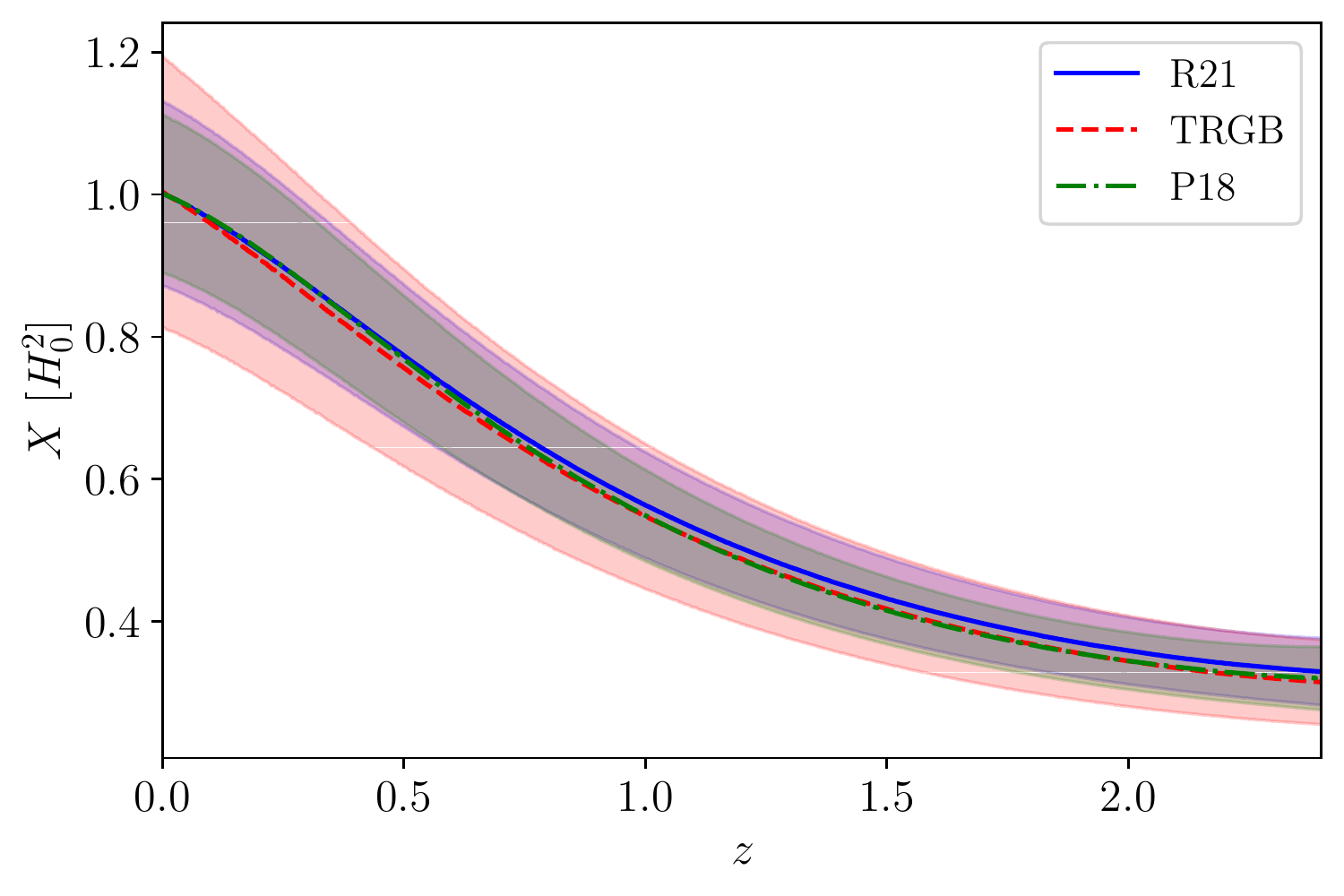}
}
\hfill
\subfigure[ ]{
\includegraphics[width = 0.475 \textwidth]{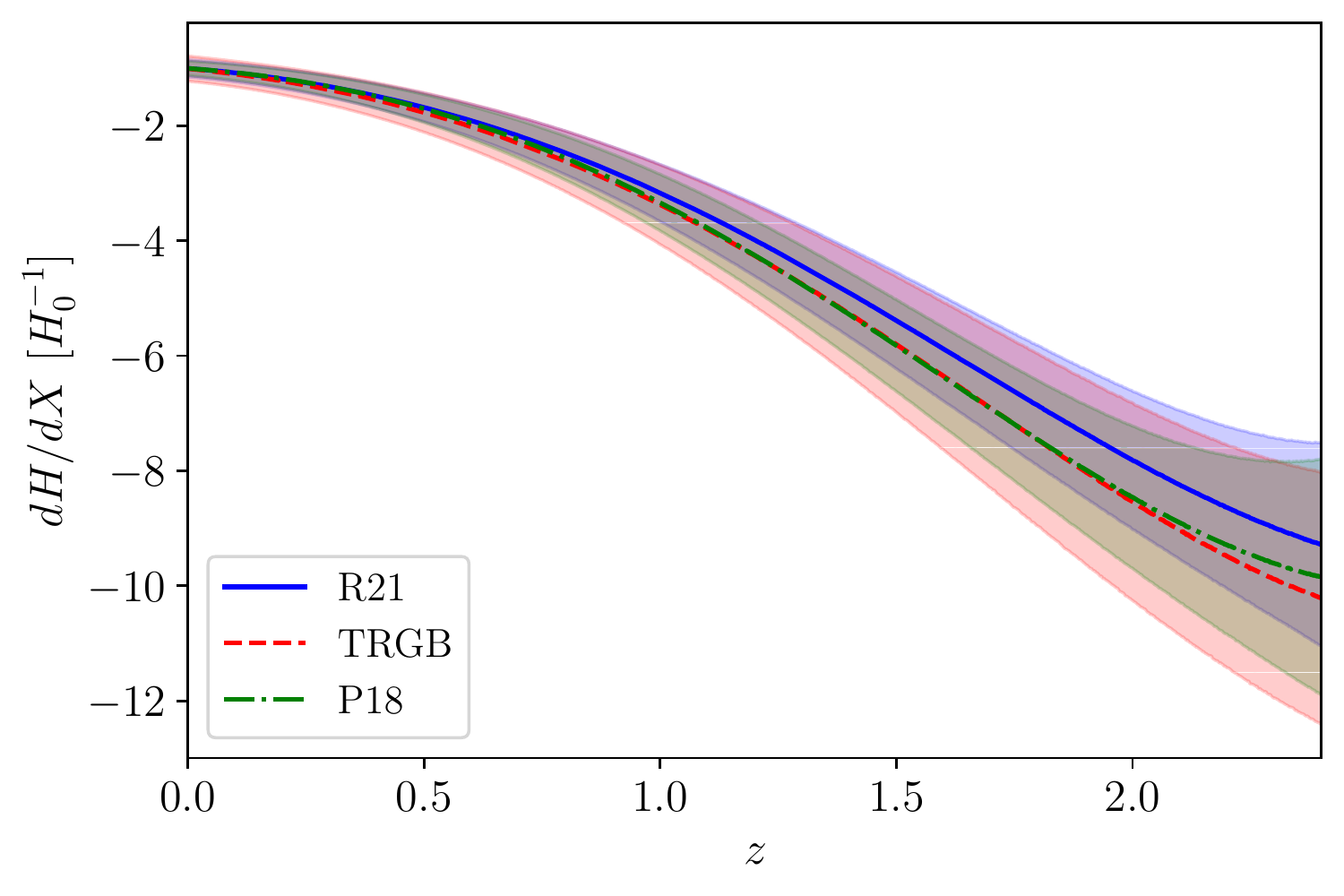}
}
\subfigure[ ]{
\includegraphics[width = 0.475 \textwidth]{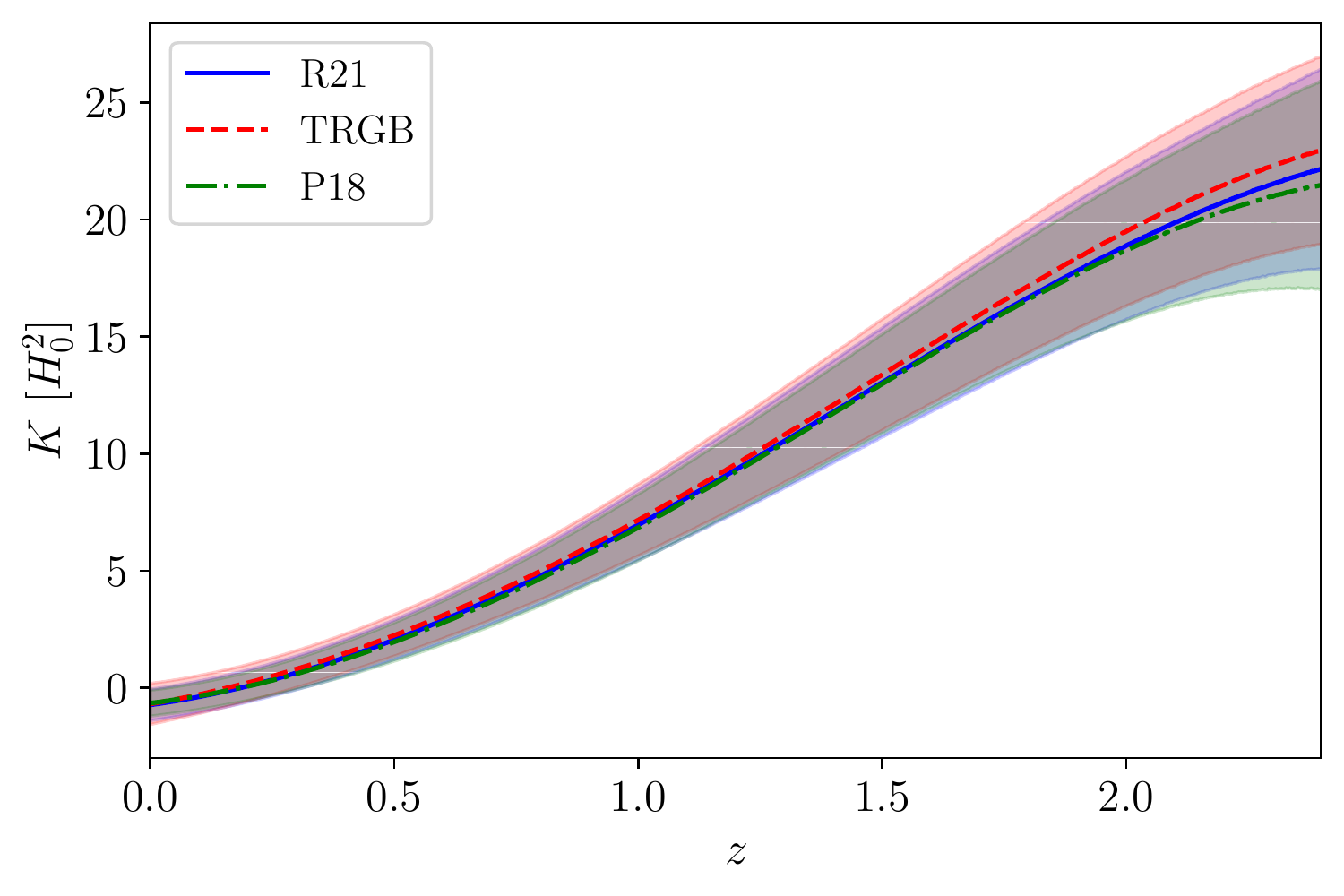}
}
\hfill
\subfigure[ ]{
\includegraphics[width = 0.475 \textwidth]{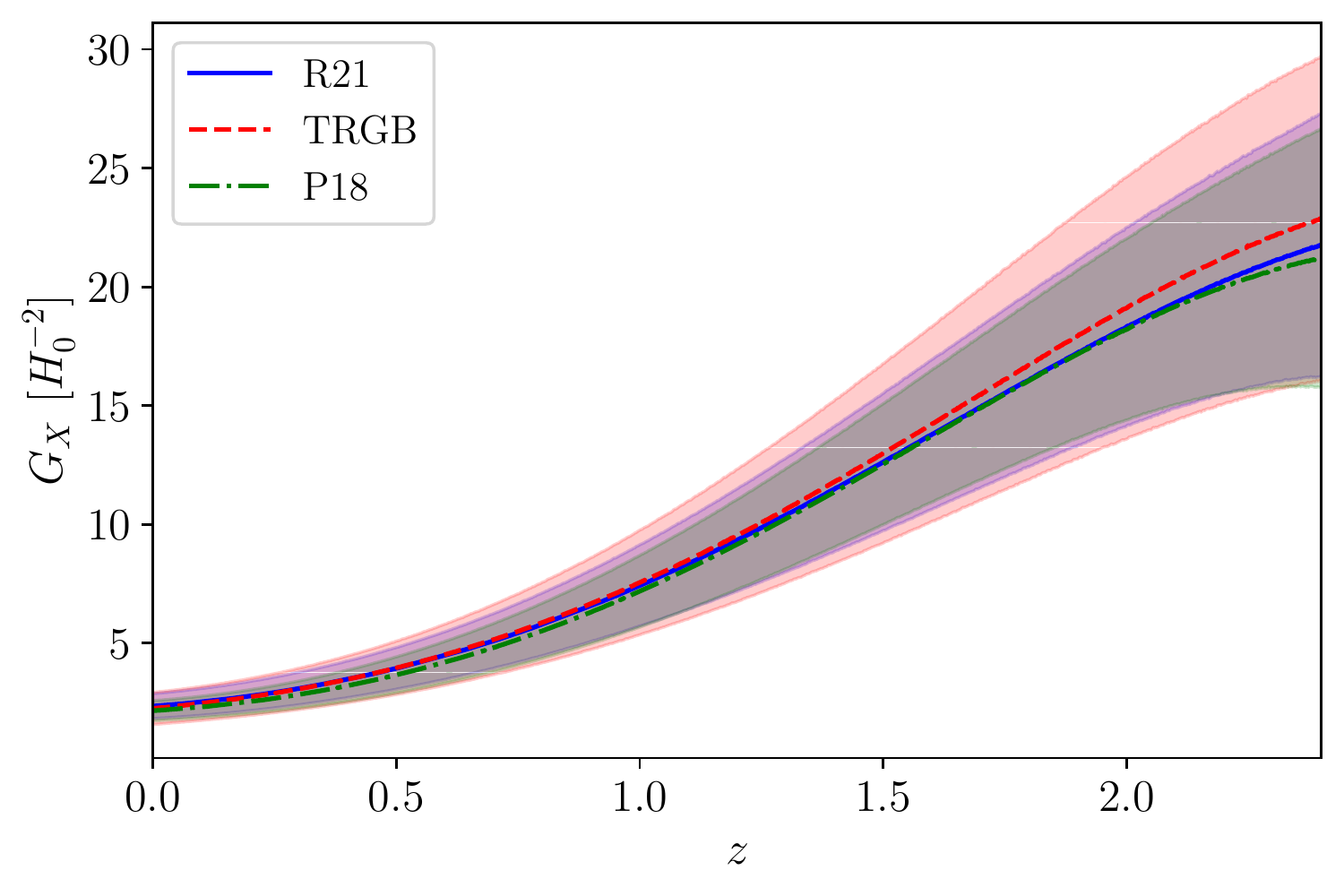}
}
\hfill
\subfigure[ ]{
\includegraphics[width = 0.475 \textwidth]{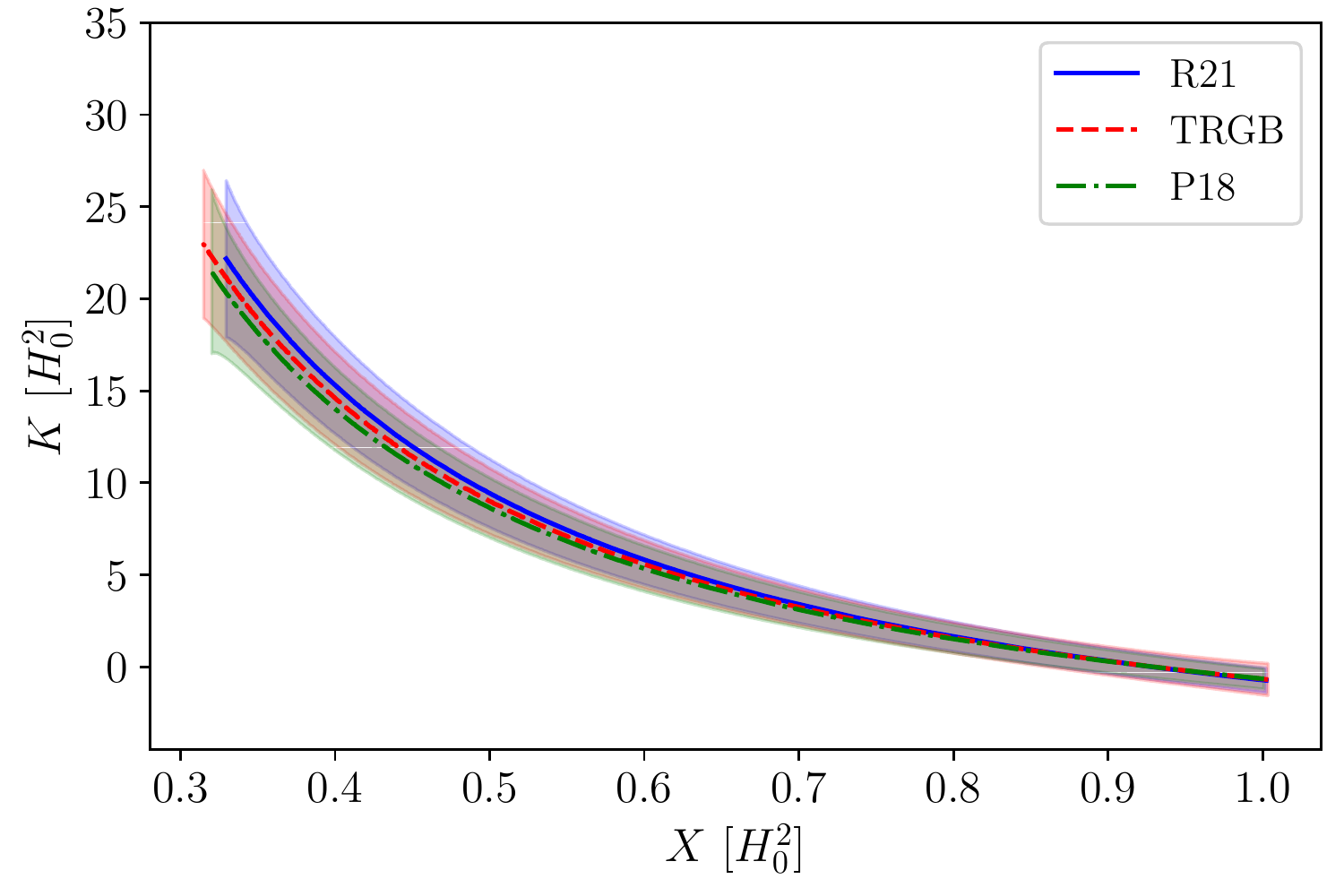}
}
\hfill
\subfigure[ ]{
\includegraphics[width = 0.475 \textwidth]{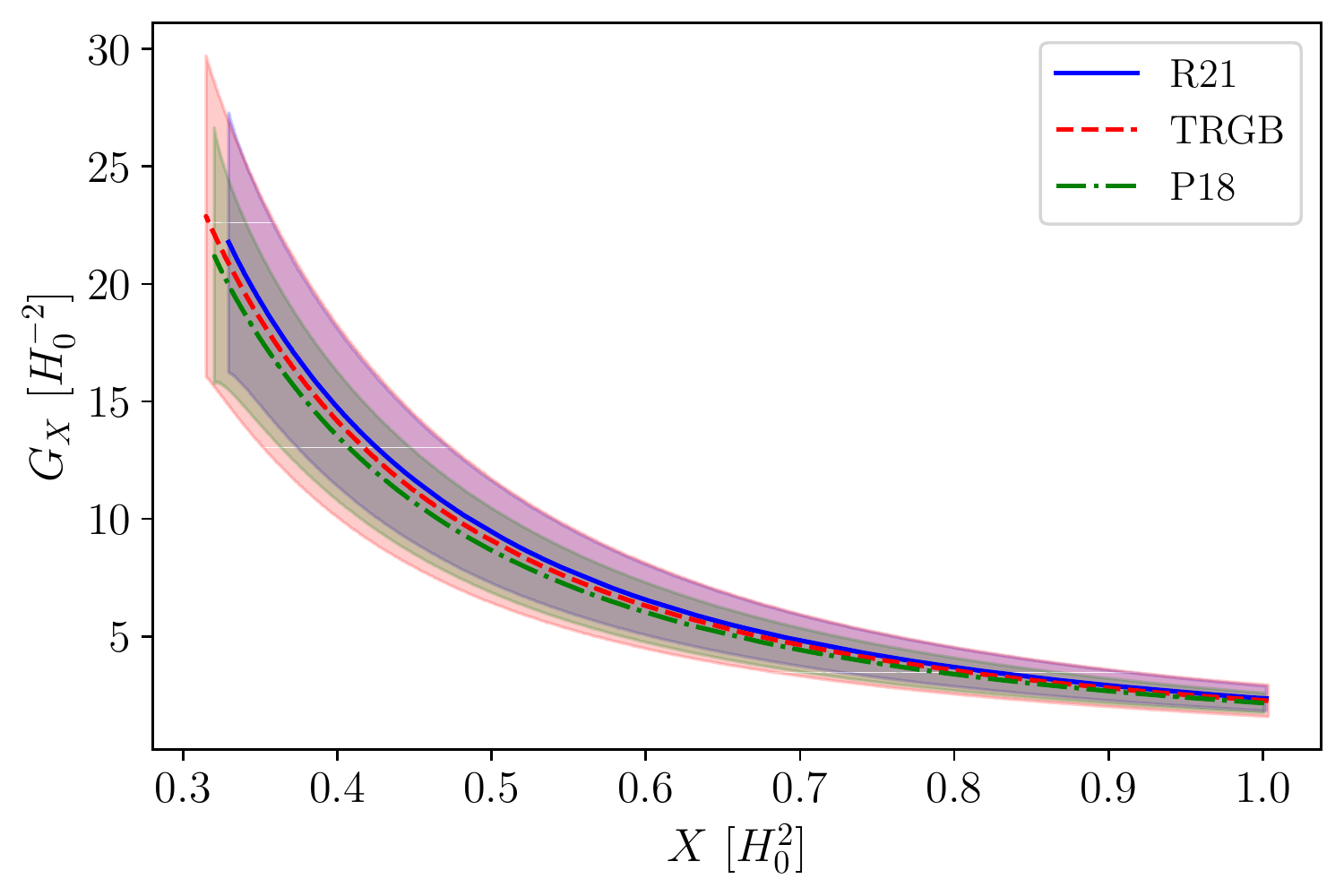}
}
\caption{Plots for the reconstructed designer Horndeski model functions (a) $X(z)$, (b) $dH/dX(z)$, (c) $K(z)$, (d) $K(X)$, (e) $G_X(z)$ and (f) $G_X(X)$, using neural networks considering R21, TRGB, and P18 $H_0$ priors. The shaded regions with `$-$', `$\vert$' and `$\times$' hatches represent the 2$\sigma$ confidence levels for the R21, TRGB, and P18 $H_0$ priors respectively.}
\label{fig:designer}
\end{figure}

Finally, with the reconstructed functions, $H(z)$ and $H'(z)$, we obtain the predictions for $X$ and $H'(X)$, as well as the designer Horndeski potentials, $K(X)$ and $G(X)$, shown in Fig. \ref{fig:designer}. In all the cases, we find that the mean $z = 0$ prediction for each $H_0$ prior is always within the $2\sigma$ level of the other two. The plots show that the TRGB $H_0$ prior gives less-constrained results for the given redshift range when compared to the R21 and P18 $H_0$ priors.

\begin{figure}[!t]
\centering
\includegraphics[width=0.485\textwidth]{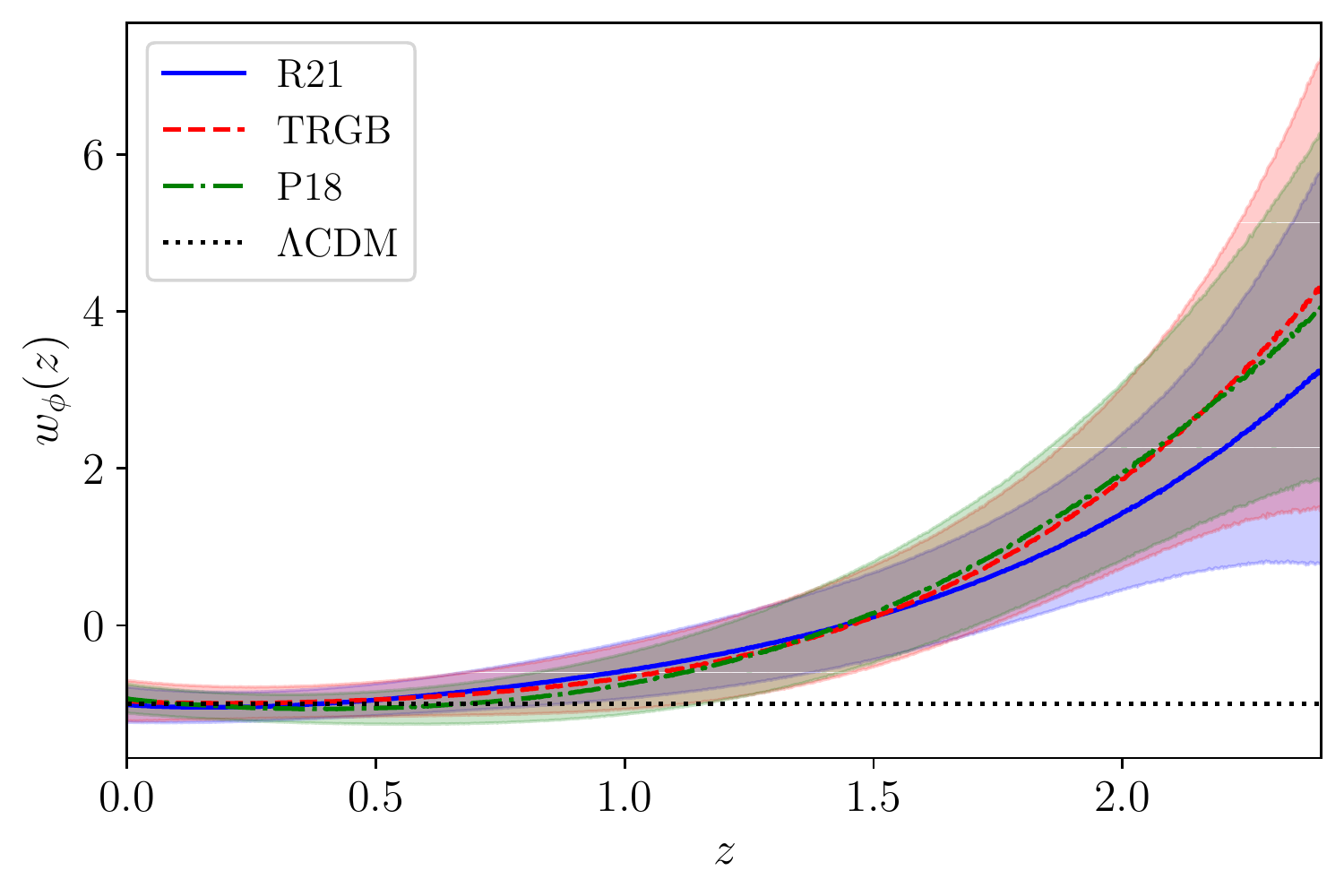} \hfill \includegraphics[width=0.485\textwidth]{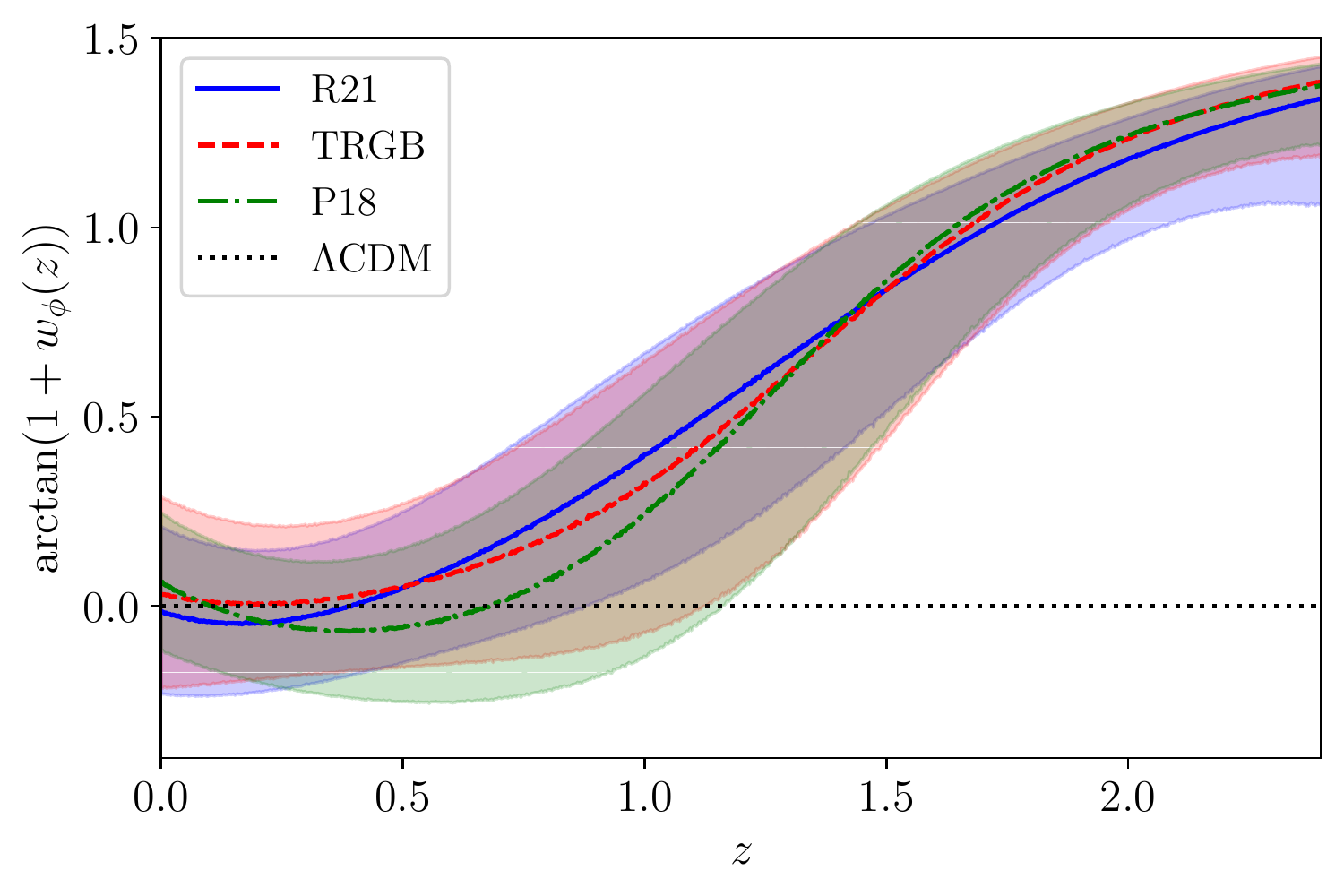}
\caption{Plots for the designer Horndeski dark energy EoS $w_\phi(z)$ (in the left panel) and its compactified form $\arctan(1+w_\phi(z))$ (in the right panel) considering R21, TRGB, and P18 $H_0$ priors. The shaded regions with `$-$', `$\vert$' and `$\times$' hatches represent the 1$\sigma$ confidence levels for the R21, TRGB, and P18 $H_0$ priors respectively.} \label{fig:w_design}
\end{figure}

\begin{figure}[!t]
\centering
\includegraphics[width=0.325\textwidth]{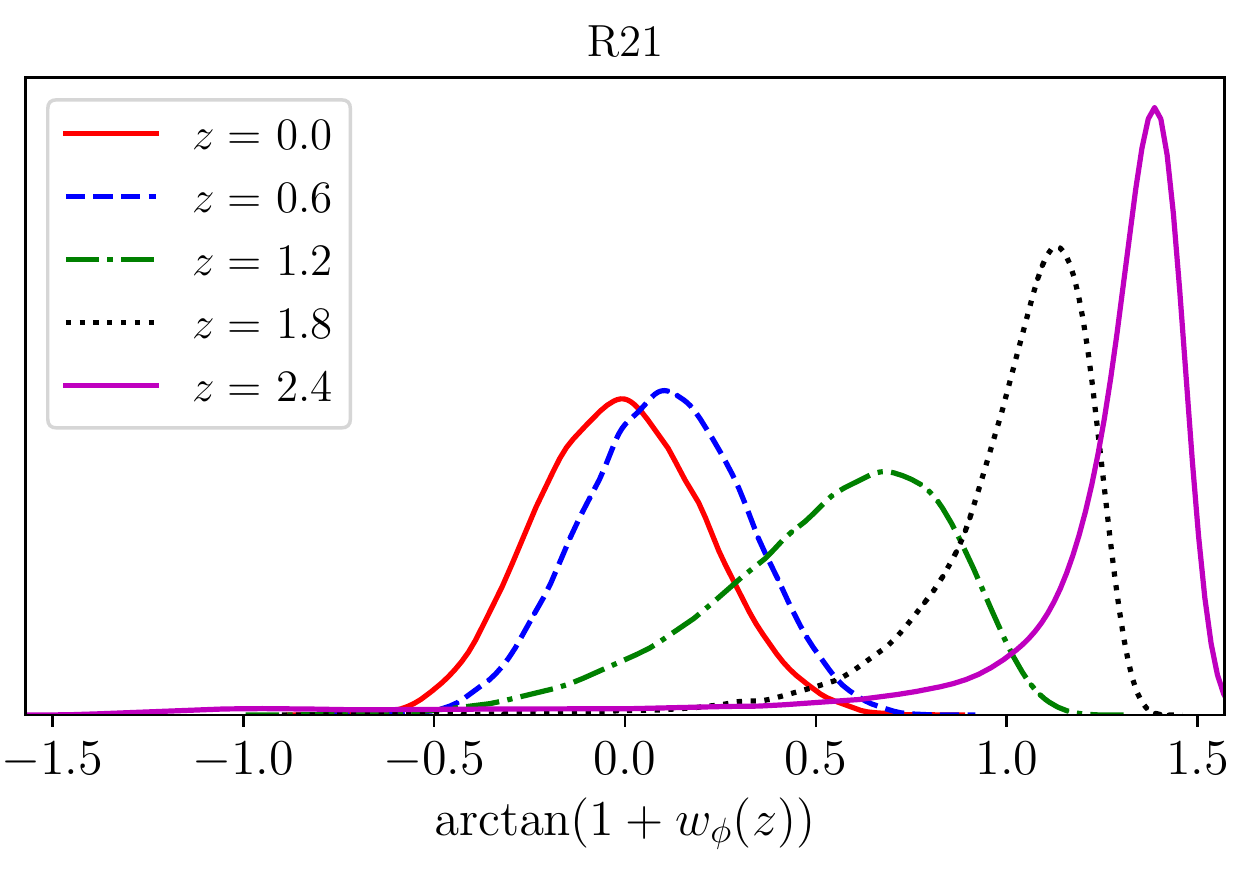} \hfill \includegraphics[width=0.325\textwidth]{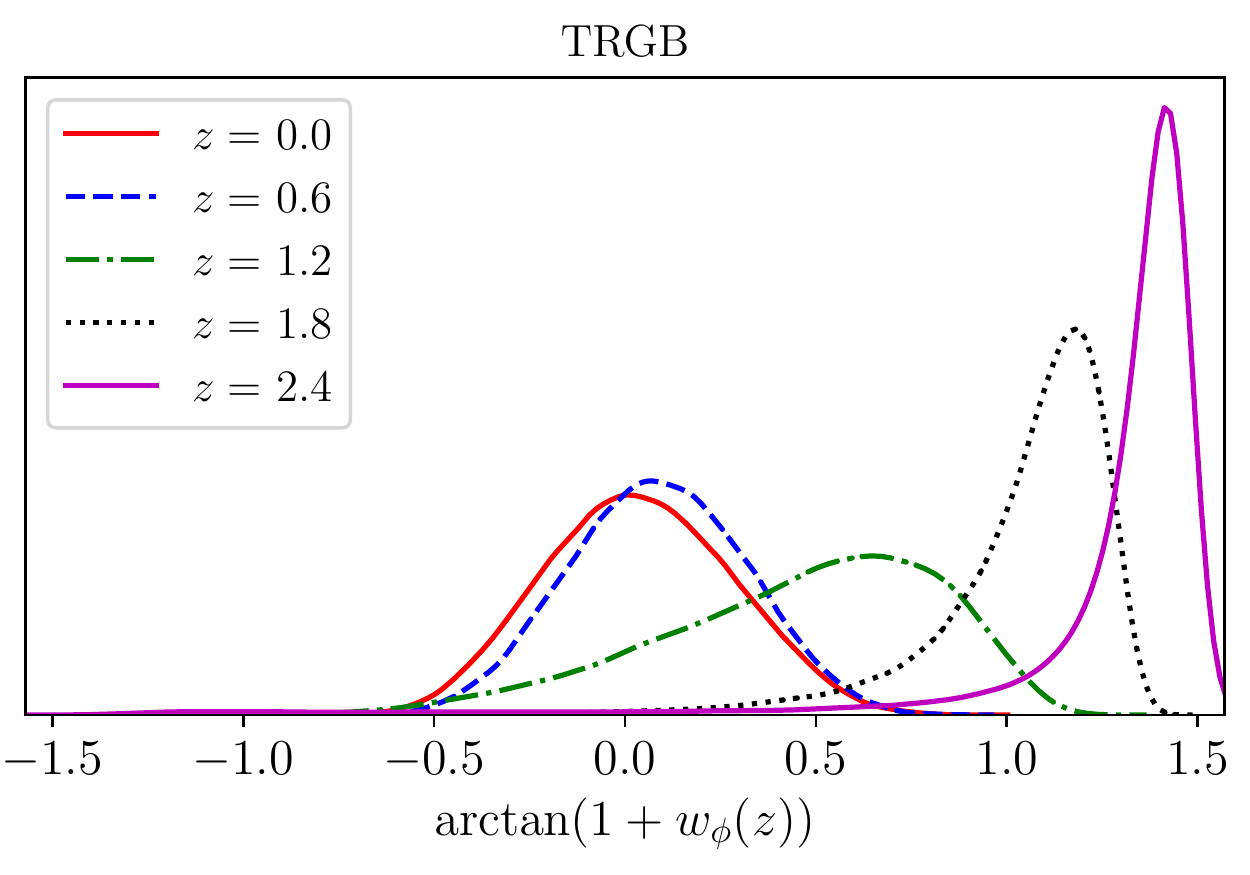} \hfill \includegraphics[width=0.325\textwidth]{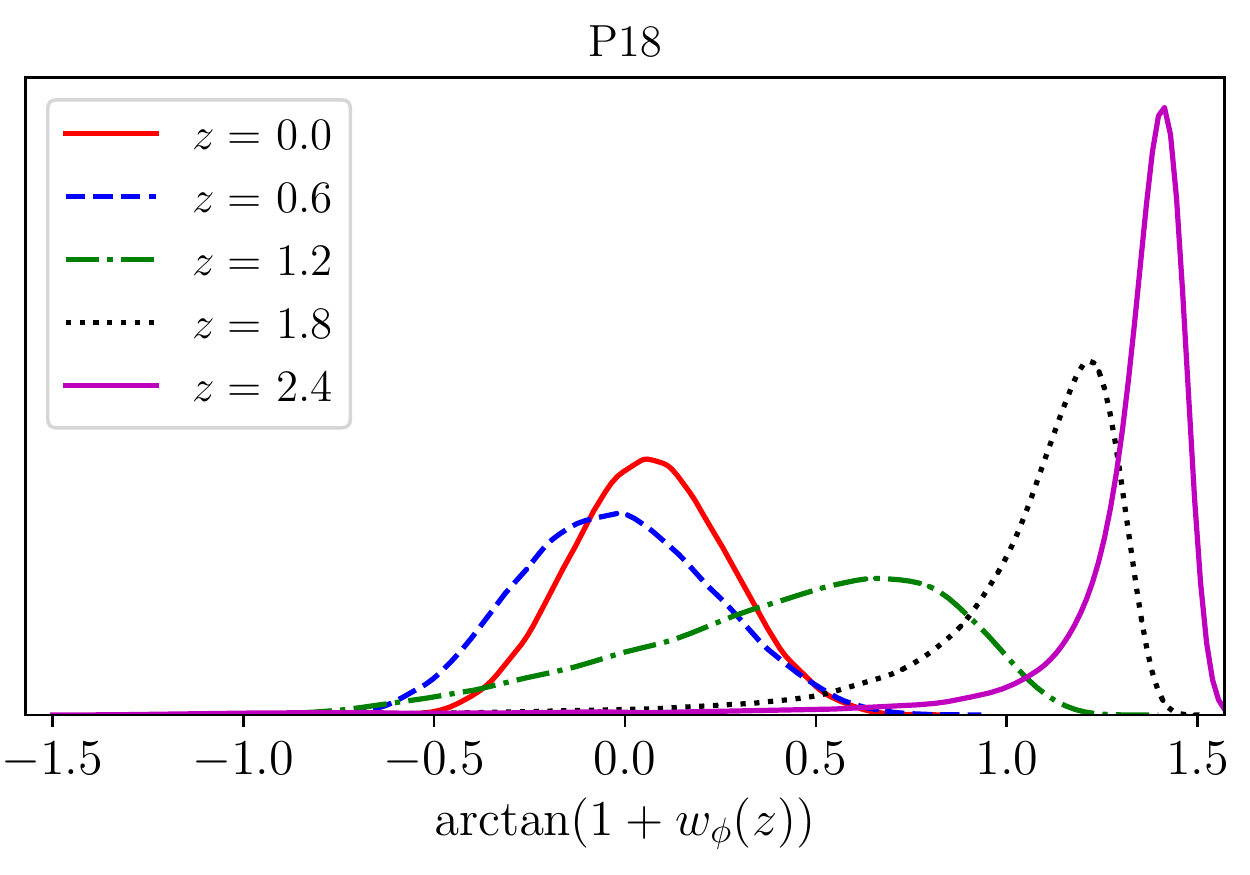}
\caption{Plots showing the posterior distribution of the compactified dark energy EoS for designer Horndeski at some sample redshifts for the R21, TRGB, and P18 H0 priors, respectively.} \label{fig:atw_dist_design}
\end{figure}

In the designer Horndeski case, the dark energy EoS takes the form,
\begin{equation}
\label{eq:w_de_hdes}
w_\phi(z) = -1 + \dfrac{ \mathcal{J} \sqrt{2X} \left( H(z)^2 - H_0^2 \left(1-\Omega_{m0}\right) \right) }{ 3 H_0^4 \Omega_{m0} \left( 1 -\Omega_{m0}\right) } - \dfrac{ 2 \mathcal{J} \sqrt{2X} (1 + z) H(z) H'(z) }{ 9 H_0^4 \Omega_{m0} \Omega_\Lambda }\,.
\end{equation}
Fig.~\ref{fig:w_design} shows the evolution of the dark energy EoS and its compactified form across redshift space for our designer Horndeski model. Interestingly enough, the compactified dark energy EoS here shows a significant difference from the respective case in quintessence, which indicates a clear deviation from $\Lambda$CDM at higher redshifts. In particular, one can notice in Fig. \ref{fig:atw_dist_design}, the $z>1$ deviations from $\Lambda$CDM ($\arctan(1+w_\phi) = 0$) for all priors, which is expected from the fact that the higher-order self interactions of the scalar field become significant at those redshifts.

\subsection{Tailoring Horndeski}

The field equations for the metric and the scalar field in tailoring Horndeski \eqref{eq:tailoring-Horndeski} are
\begin{gather}
    3H^2 = \rho + 2 \Lambda + \frac{\dot{\phi}^2}{2} + 3 H \dot{\phi}^3 G_X \,, \label{eq:tail_fried}\\
    2 \dot{H} + 3 H^2 = - P + 2 \Lambda - \frac{\dot{\phi}^2}{2} + G_X \dot{\phi}^2 \ddot{\phi}\,,\\
    \ddot{\phi} \left[ 3 H \left( G_{XX}\dot{\phi}^3 + 2 G_X \dot{\phi}\right)+1\right] + 3 \dot{\phi} \left(G_X \dot{H} \dot{\phi} + 3 G_X H^2 \dot{\phi} + H \right) = 0
\end{gather}
The solution of this system of equations will lie on the dynamical hypersurface determined by Eq. \eqref{eq:hdes_sf} with $J=0$, as a dynamical attractor, since we can get these equations by setting $K(X) = X - 2\Lambda$ in the set \eqref{eq:hdes_fried1}-\eqref{eq:hdes_fried3} of designer Horndeski. One can then write, 
\begin{equation}
G_X (X) = - \dfrac{1}{3 \sqrt{2 X} H(X)}\,.\label{eq:GX_tail}
\end{equation}
This $G_X$ can be used to obtain the following model-independent necessary conditions, 
\begin{equation}
X = 3 \left( H_0^2 \Omega_m(z) + H_0^2 \Omega_\Lambda - H^2 \right)\,, \label{eq:X_tail}
\end{equation}
and
\begin{equation}
2 \dot{H} + 3 H^2 = -P + 3 H_0^2 \Omega_\Lambda - X - \dfrac{\dot{X}}{3 H}\,, \label{eq:Hdot_tail}
\end{equation}
where $\Omega _m (z) = \rho (z)/(3H_0)^2$ and $\Omega _\Lambda = 2 \Lambda /(3H_0)^2$.
This means that for a given Hubble function, one can uniquely determine the kinetic term of the scalar field using Eq.~\eqref{eq:X_tail}. 
Therefore, the potential $G(X)$ and the scalar field's kinetic term can be assigned to a given $H$ through Eqs.~\eqref{eq:GX_tail} and \eqref{eq:X_tail} respectively. 

The dark energy EoS for tailoring Horndeski is given by
\begin{equation}
w_\phi(z) = \dfrac{ H(z) \left( 3H(z) - 2 ( 1 + z ) H'(z) \right) }{ 3 \left( \Omega_{m0} H_0^2 (1 + z)^3 -H^2(z) \right)  } ,
\end{equation}
which, is the exact same expression obtained for the quintessence dark energy EoS in Eq. \eqref{eq:w_de_quint} and also can be obtained from designer Horndeski by setting $K(X) = X - 2\Lambda$.

\begin{figure}[!t]
\centering
\includegraphics[width = 0.325\textwidth]{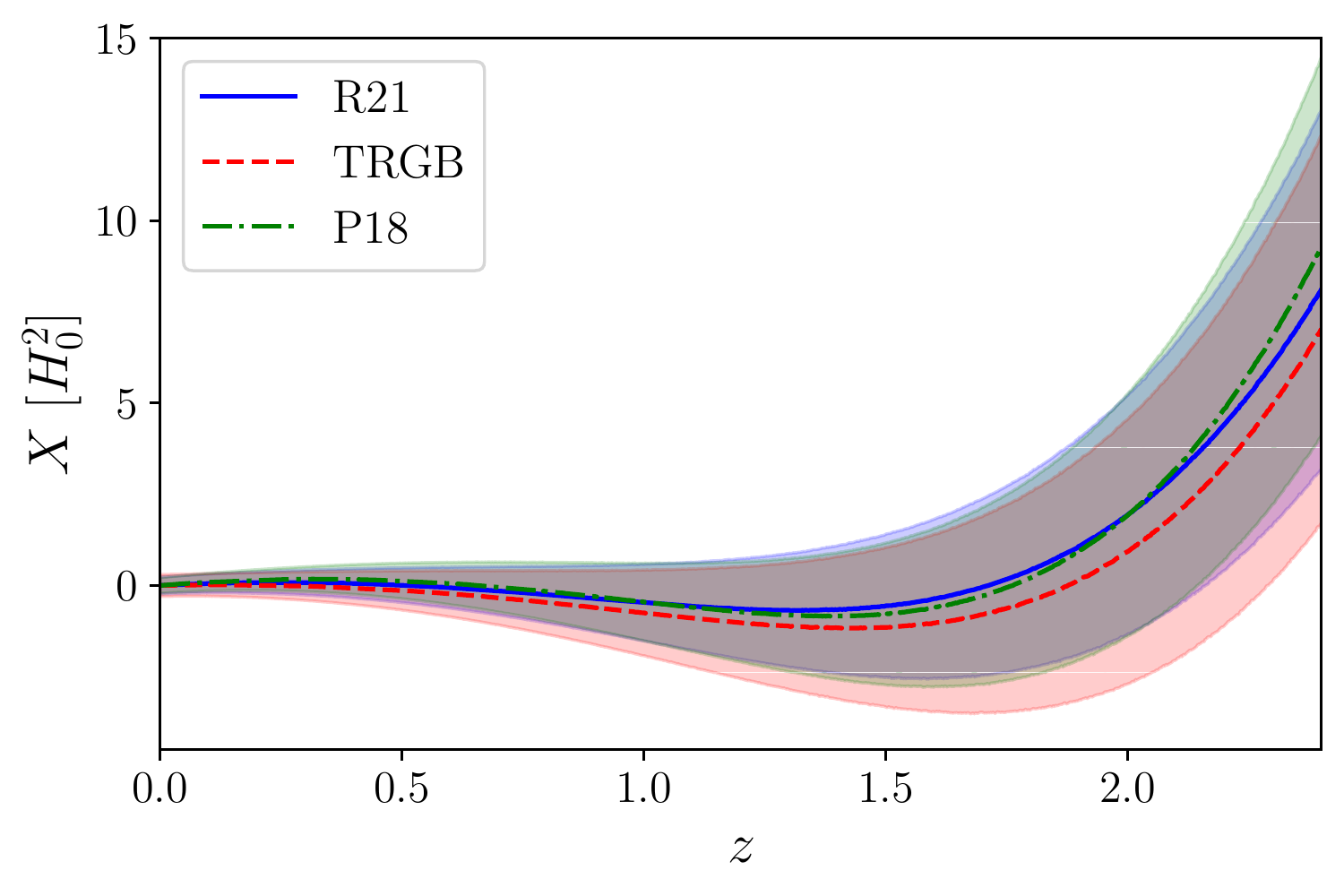}
\hfill
\includegraphics[width = 0.325\textwidth]{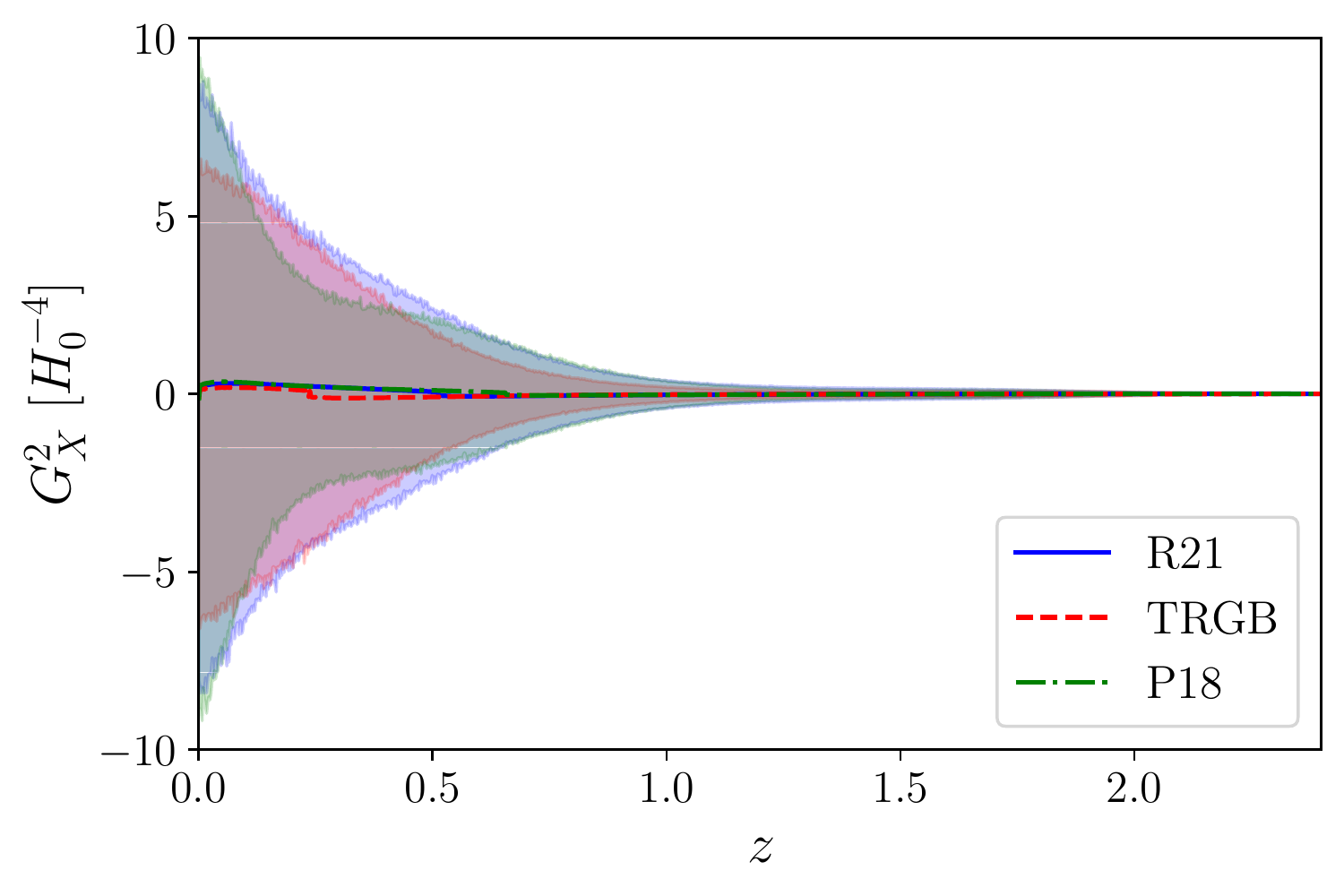}
\hfill
\includegraphics[width = 0.325\textwidth]{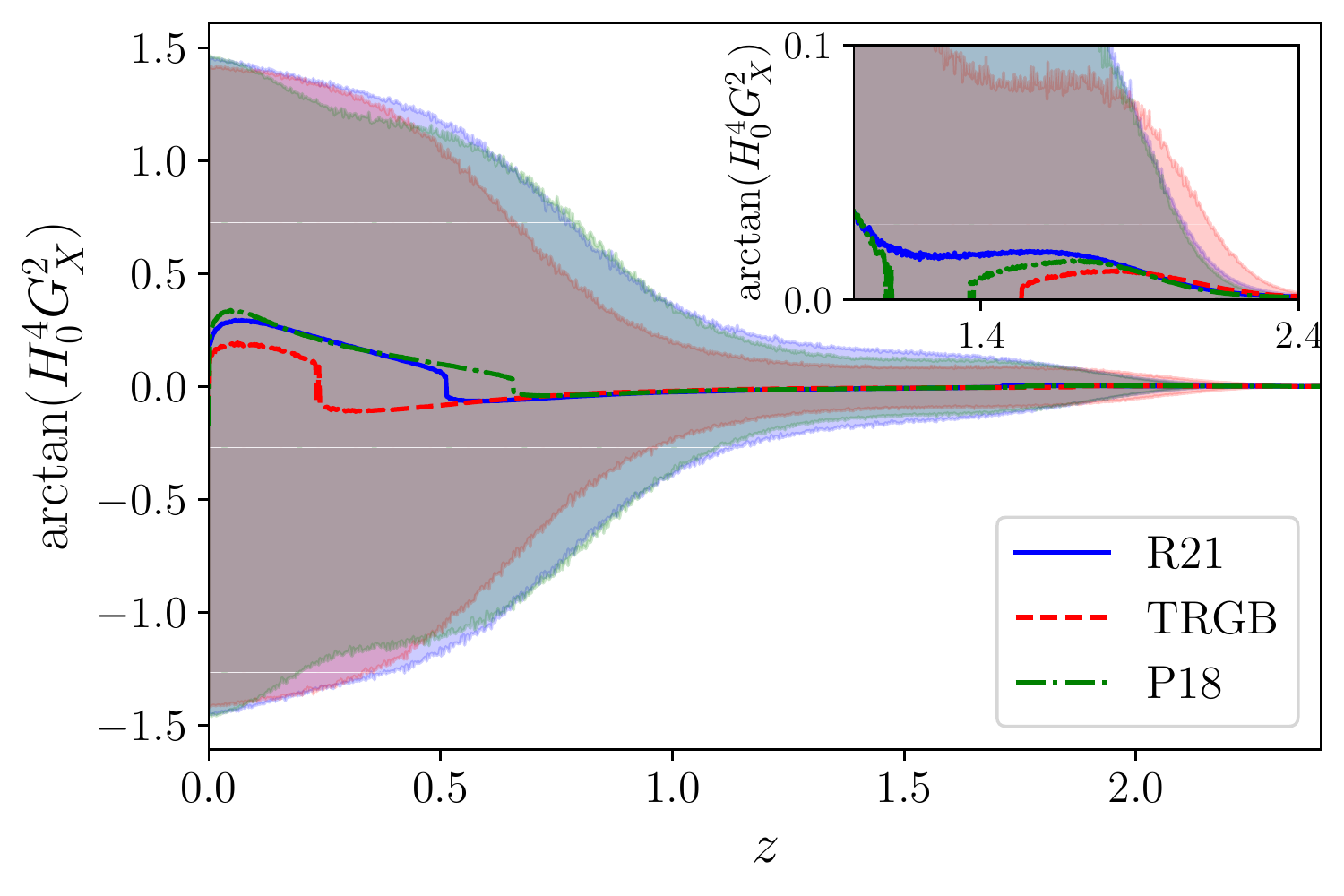}
\caption{Plots for the reconstructed tailoring Horndeski model functions $X$ (left panel), $G_X^2$ (middle panel), and $\tilde{G}_X^2$ (right panel), using neural networks considering R21, TRGB, and P18 $H_0$ priors. The shaded regions with `$-$', `$\vert$' and `$\times$' hatches represent the 2$\sigma$ confidence levels for the R21, TRGB, and P18 $H_0$ priors respectively.}
\label{fig:tailoring}
\end{figure}

\begin{figure}[!t]
\centering
\includegraphics[width = 0.325\textwidth]{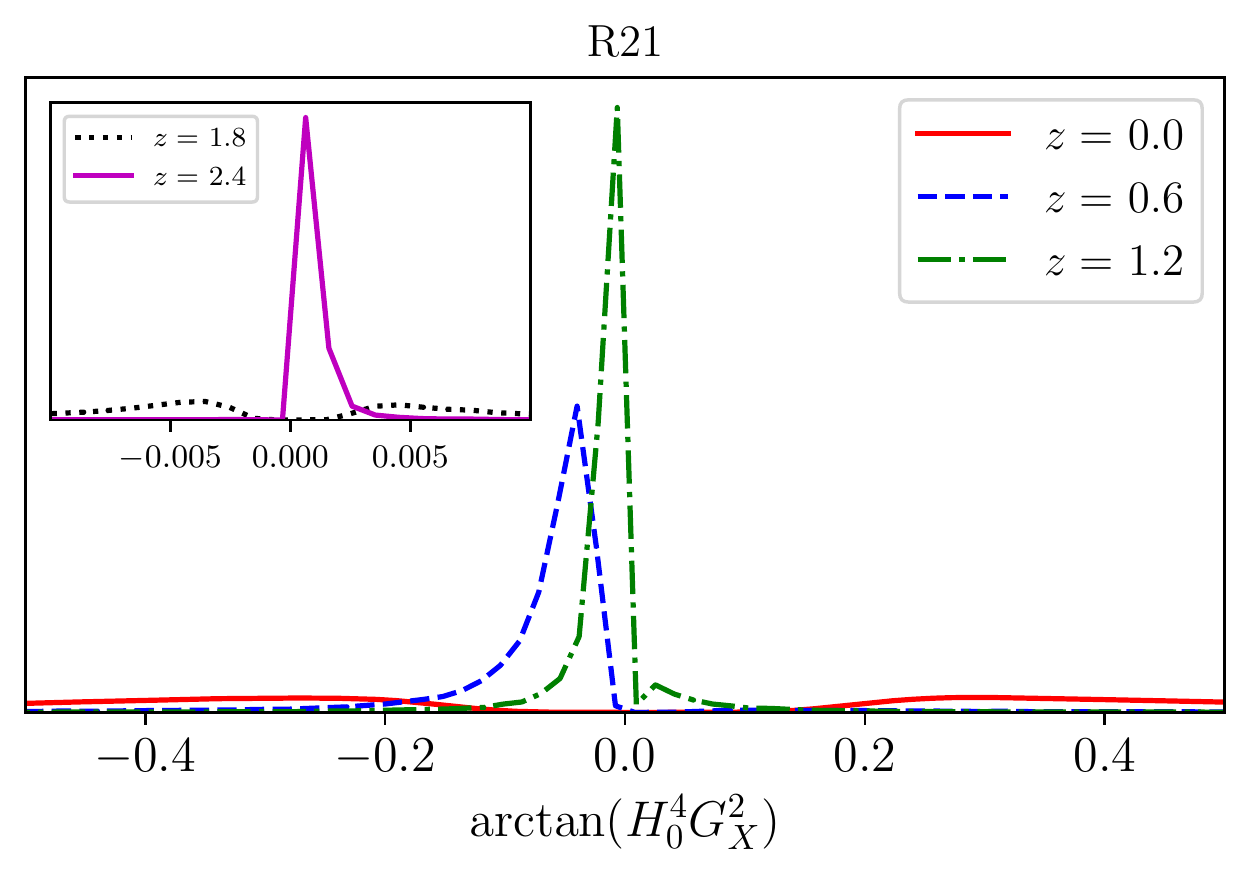}
\hfill
\includegraphics[width = 0.325\textwidth]{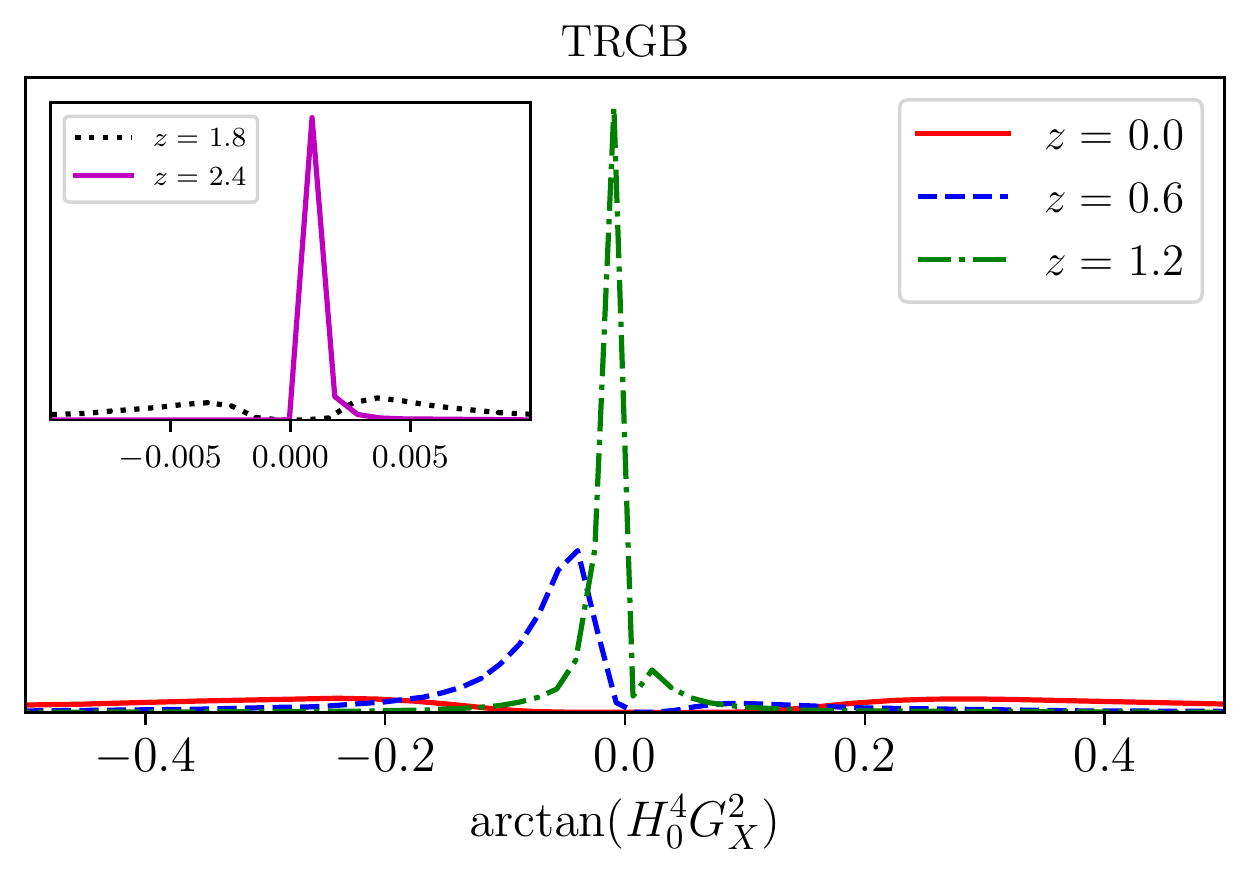}
\hfill
\includegraphics[width = 0.325\textwidth]{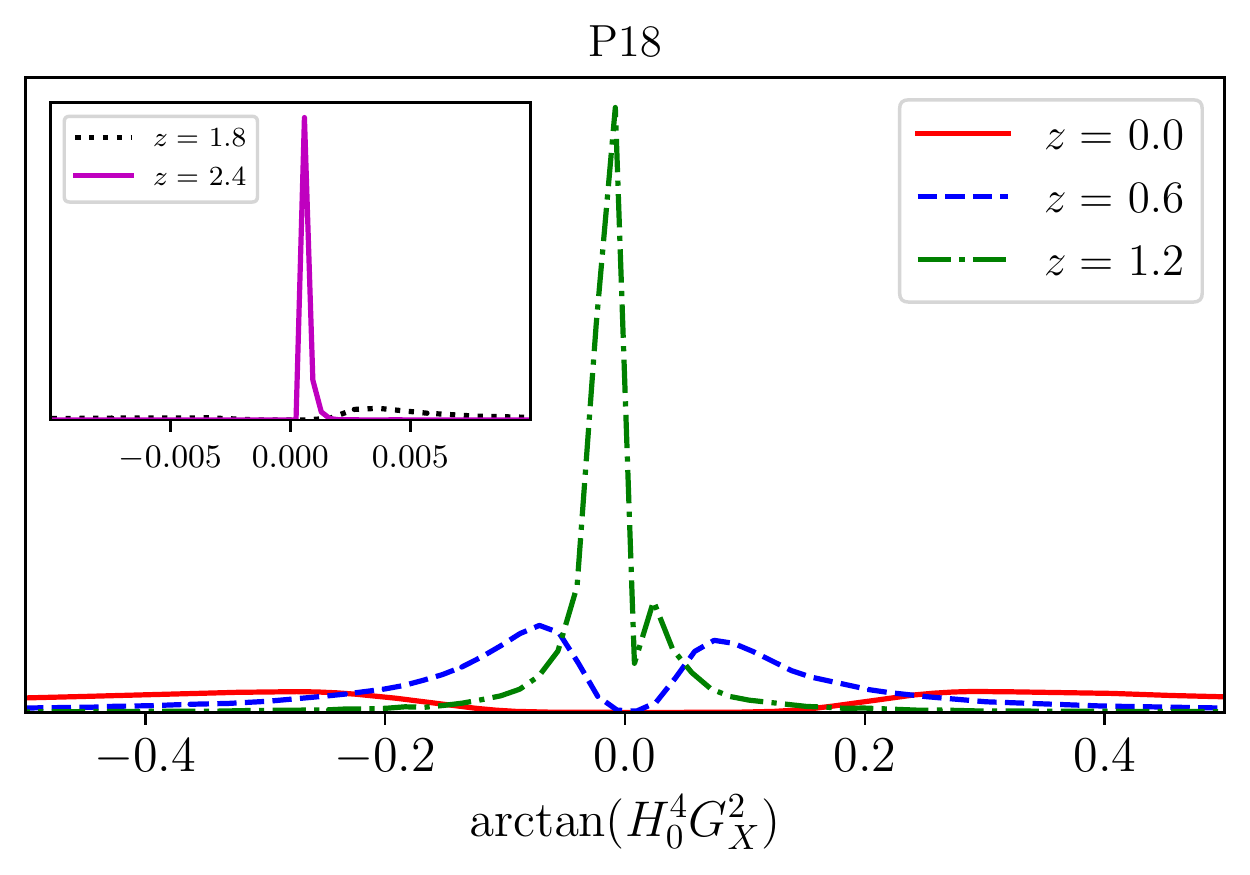}
\caption{Plots showing the posterior distribution of the compactified braiding potential, $\tilde{G}_X^2$, for the tailoring Horndeski model at some sample redshifts for the R21, TRGB, and P18 H0 priors, respectively.} \label{fig:atGX_dist_tailor}
\end{figure}

Using the reconstructed functions $H(z)$ and $H'(z)$, we can plot the evolution of $X$ and $G_X$, as shown in Fig. \ref{fig:tailoring}. Additionally, we sample over another compactified variable, $\tilde{G}_X^2$, referred to as the compactified braiding potential,  
\begin{equation}\label{eq:compactified_braiding}
    \tilde{G}_X^2 = \arctan \left( G_X^2 H_0^4 \right) = \arctan\left( \frac{H_0^4}{18 X H\left(X \right)^2} \right)\,.
\end{equation}
The plot for this compactified braiding potential and its associated posterior distribution, at different redshifts, is shown in Fig. \ref{fig:atGX_dist_tailor}.

We further summarize the obtained constraints on the dark energy EoS at $z = 0$ for the Horndeski models in Table~\ref{tab:w_de_constraints}. One can observe that for current times, all the analyses produce an EoS that is equal to $w_{\rm \Lambda CDM} = -1$ to within $1\sigma$. On the other hand, there are some important nuances, the Quintessence \& Tailoring Horndeski models have a lower EoS only for the case in which the $H_{0}^{\rm R21}$ prior is assumed while it expresses phantom behaviour in the other cases. The opposite situaiton occurs for the Designer Horndeski scenario. Indeed, for the $H_{0}^{\rm R21}$ prior one find a high value of the Hubble constant best fit while differing values of the EoS. This gives the models distinct flavors that are important.

\begin{table}
\caption{Constraints on the dark energy EoS in Horndeski cosmology, obtained via reconstruction using corresponding to the R21, TRGB and P18 $H_0$ priors.}
\resizebox{\textwidth}{!}{\renewcommand{\arraystretch}{1.5} \setlength{\tabcolsep}{20 pt} \centering
\begin{tabular}{ c c c c c }
\hline
Models & Hubble prior & $H_0$ & $\Omega_{m0}$ & $w_\phi(z=0)$ \\
\hline \hline
\textit{Quintessence}  & R21 & $73.3 \pm 1.04$ &  $0.286 \pm 0.018$ & $-0.992^{+0.126}_{-0.130}$ \\
 \& & TRGB & $69.8 \pm 1.7$ & $0.301 \pm 0.019$ &  $-1.012^{+0.139}_{-0.142}$ \\
\textit{Tailoring Horndeski}  & P18 & $67.4 \pm 0.5$ & $0.311 \pm 0.017$ & $-1.035^{+0.114}_{-0.117}$ \\

\hline

  & R21 & $73.3 \pm 1.04$ &  $0.286 \pm 0.018$ &  $-1.013^{+0.237}_{-0.235}$\\
\textit{Designer Horndeski} & TRGB & $69.8 \pm 1.7$ & $0.301 \pm 0.019$ & $-0.988^{+0.252}_{-0.265}$ \\
  & P18 & $67.4 \pm 0.5$ & $0.311 \pm 0.017$ &  $-0.941^{+0.207}_{-0.203}$\\

\hline

\hline
\end{tabular}
}
\label{tab:w_de_constraints}
\end{table}


\section{Conclusion \label{sec:conclusion}}

In this work, we have shown how a native neural network implementation for cosmological data sets could be used to select physical models from general classes of scalar-tensor theories from Horndeski gravity. Horndeski gravity offers an ideal framework in which to construct physical models of cosmology involving the evolution of a scalar field that interacts with the gravitational sector to produce changes in the evolution of cosmological parameters. The observational constraints on possible models coming from gravitational wave observations continue to allow a vast range of possible models which host a scalar field but which may have a multitude of potential forms. For this reason, more sophisticated model discrimination tools are necessary to restrict the space of viable models allowed by observational data sets. This work is one possible in that direct, and one potential implementation that can limit the specific forms of potential Horndeski models.

In our implementation, we used late time data to train the various ANNs through the lens of CC, SN, and BAO data sets, as well as the inclusion of select priors on the Hubble constant. One of the novel features of this work is that we incorporate the full complexity of the data sets, including covariance matrices, into our ANN training regime. This is an important point since this likelihood-free approach to reconstructing the evolution profiles of cosmic parameters requires a new approach as compared with traditional MCMC likelihood functions, where it is well known how to incorporate this information. In our case, we take the route of putting the covariances into the loss function definitions, so that the complexity of the data is naturally integrated into the analysis at the level of the training of the ANNs.

In our study, we consider Quintessence, designer, and Tailoring Horndeski models. In our first iteration, we consider a minimally coupled scalar field which is identical to a Quintessence field~\eqref{eq:quintessence} with an EoS that depends dynamically on the evolution of the scalar field. Here, we find reconstructions of the model potential and scalar field derivative (which is related to the kinetic term) functions in Fig.~\ref{fig:quint}, in which a preference is found for a gradually increasing potential and a kinetic term that does deviate from $\Lambda$CDM at $2\sigma$ for higher redshifts. On the other hand, in Figs.~\ref{fig:w_quint} and \ref{fig:atw_dist_quint}, the EoS for this scalar field does agree with current constraints of $-1$ at very low redshifts but can feature evolution profiles at other points. In Fig.~\ref{fig:atw_dist_quint} the preferred EoS value is significantly different from this value.

The Designer Horndeski model~\eqref{eq:designer-Horndeski} contains more complex terms in the Horndeski components. Of particular importance is the appearance of higher derivative scalar field terms which may be impactful for parts of the background evolution profiles for the background parameters. Despite the model needing more functions in order to be prescribed, our approach continues to give good constraints at redshift ranges of interest, as evidenced in Fig.~\ref{fig:designer}. Here, the reconstruction approach tends to prefer model behavior that gives more variance to the kinetic term which is balanced by the other functional behaviors. In difference to the Quintessence model, the scalar field EoS now does not feature a minimum point and rises monotonically with redshift, indicating a tendency to this component to be less exotic at earlier times, as shown in Fig.~\ref{fig:w_design}. This is compounded by the distribution of the EoS values at different redshift points in Fig.~\ref{fig:atw_dist_design}. On the other hand, the Tailoring Horndeski model gives an altogether different evolution with a scalar field kinetic term that grows with redshift, as shown in Fig.~\ref{fig:tailoring}, while the EoS is identical to that of the Quintessence model. 

It would be interesting to take this initial work on the topic and consider observations related to the perturbative part of the theory such as using measurements of large scale structure or mock data related to cosmological gravitational wave observations, in the context of ANNs. We hope to investigate these open questions in future work on this topic.

\acknowledgments

This paper is based upon work from COST Action CA21136 {\it Addressing observational tensions in cosmology with systematics and fundamental physics} (CosmoVerse) supported by COST (European Cooperation in Science and Technology). The work was supported by the PNRR-III-C9-2022–I9 call, with project number 760016/27.01.2023 and by the Hellenic Foundation for Research and Innovation (H.F.R.I.) under the ''First Call for H.F.R.I. Research Projects to support Faculty members and Researchers and the procurement of high-cost research equipment grant'' (Project Number: 2251).

\appendix


\bibliographystyle{JHEP}
\bibliography{references}

\label{lastpage}

\end{document}